\documentclass[letterpaper,english,reprint, aps, prx]{revtex4-1}
\usepackage[T1]{fontenc}
\usepackage[utf8]{inputenc}
\usepackage{verbatim}
\usepackage{amsmath}
\usepackage{amssymb}

\makeatletter

\usepackage{graphicx}
\usepackage{xcolor}
\usepackage[bookmarks=false,linkcolor=blue,urlcolor=blue,colorlinks,citecolor=blue]{hyperref}

\makeatother

\usepackage{babel}
\begin{document}

\preprint{}

\title{Variational-Correlations Approach to Quantum Many-body Problems}

\author{Arbel Haim,$^{1,2}$ Richard Kueng,$^{1,3}$ Gil Refael$^{1}$}

\affiliation{$^1$Institute for Quantum Information and Matter and Department of Physics, California Institute of Technology, Pasadena, CA 91125, USA\\
\mbox{$^2$Walter Burke Institute for Theoretical Physics, California Institute of Technology, Pasadena, CA 91125, USA}\\
\mbox{$^3$Department of Computing and Mathematical Sciences, California Institute of Technology, Pasadena, CA 91125, USA}}

\date{\today}
\begin{abstract}
We investigate an approach for studying the ground state of a quantum
many-body Hamiltonian that is based on treating the correlation functions
as variational parameters. In this approach, the challenge set by
the exponentially-large Hilbert space is circumvented by approximating
the positivity of the density matrix, order-by-order, in a way that
keeps track of a limited set of correlation functions. In particular,
the density-matrix description is replaced by a correlation matrix
whose dimension is kept linear in system size, to all orders of the
approximation. Unlike the conventional variational principle which
provides an upper bound on the ground-state energy, in this approach
one obtains a lower bound instead. By treating several one-dimensional
spin $1/2$ Hamiltonians, we demonstrate the ability of this approach
to produce long-range correlations, and a ground-state energy that
converges to the exact result. Possible extensions, including to higher-excited
states are discussed.
\end{abstract}
\maketitle

\section{Introduction\label{sec:Intro}}

Systems comprising of many interacting quantum particles are encountered
in various fields, from condensed-matter and cold-atoms systems to
quantum chemistry and nuclear matter. The ability to analyze quantum
many-body systems, however, is severely limited by the exponential
amount of information needed to describe the quantum wave-function.
The challenge in studying quantum many-body systems is, therefore,
to access the relevant physical observables without having to store
and manipulate the full wave function

In one dimension (1d), the Density Matrix Renormalization Group (DMRG)
method does that by employing the Matrix Product State representation
that can describe ground states using an amount of information that
scales only as a power law with the system size (and for gapped ground
states strictly linear)~\citep{White1992PRL,White1993PRB,Schollwock2005TheDensity}.
This is not the case, however, for higher-dimensional systems or for
highly-excited states. Quantum Monte Carlo simulations~\citep{Gubernatis2016quantum}
are not limited to 1d. However, they are only suitable for systems
not suffering from the notorious ``sign problem''~\citep{Loh1990sign},
leaving out many interesting physical systems. Recently, promising
results have been achieved by employing machine learning techniques
to study quantum many-body systems~\citep{Carleo2017solving,Melko2019restricted,Sharir2019deep},
and the full potential of these methods is yet to be discovered.

In this paper, we discuss a method for numerically studying the ground
state of quantum many-body systems. This method relies on directly
accessing a limited amount of physical information, involving the
energy and several correlation functions, instead of treating the
full quantum many-body wave function. This is done by treating these
correlation functions as variational parameters in the minimization
of the ground-state energy. Importantly, constraints are placed on
the correlation functions, in a way that approximates the condition
of the density-matrix being positive semidefinite. By gradually keeping
more correlation functions, this approximation becomes increasingly
better, and the resulting ground-state energy approaches its exact
value.

Our approach follows a similar logic to that of the variational two-electron
reduced density matrix (2-RDM) method~\citep{Mazziotti2002variational,Zhao2004reduced,Mazziotti2005variational,Mazziotti2006quantum,Mazziotti2007reduced,Barthel2012Solving,Baumgratz2012Lower,Anderson2013second,Verstichel2013Extensive,Mazziotti2016enhanced,Alcoba2018direct,Rubio2019variational},
developed in the context of quantum chemistry. As in these past works,
the method we discuss involves a \emph{relaxation} of the constraints
on the many-body wave-function, and it therefore yields a lower bound
on the ground-state energy. In the present work, an emphasis is put
on limiting the amount of information kept in a way that enables applying
the approximation order by order, without changing the scaling of
the computation with the system size. This allows for the treatment
of large systems, including in the absence of translational invariance.

More specifically, this relaxation is achieved by substituting the
density-matrix description of the system, which requires an exponentially-large
amount of information, with a physical correlation matrix whose dimension
is only linear in the system size. To demonstrate the variational-correlations
approach, we treat several one-dimensional spin-1/2 models, with and
without disorder, and show the method can achieve convergence towards
the exact ground-state energy, as well as to produce long-range spin
correlations. 

\section{The variational correlations Approach\label{sec:method}}

For simplicity, we present the formalism as it applies to spin-1/2
chains with nearest neighbor interactions. An extension to more general
systems and higher dimensions is straightforward. The Hamiltonian
for such a system is most generally given by 
\begin{equation}
\hat{H}=\sum_{n=1}^{N}\sum_{i\in\{x,y,z\}}\sum_{\alpha\in\{0,x,y,z\}}J_{ni\alpha}\hat{\sigma}_{n}^{i}\hat{\sigma}_{n+1}^{\alpha},\label{eq:Hamiltonian}
\end{equation}
where $N$ is the number of spins, $n$ runs over the chain's sites,
$\hat{\sigma}_{n}^{x,y,z}$ are the Pauli matrices on site $n$, and
$\hat{\sigma}_{n}^{0}$ is the identity matrix on site $n$. The system
is assumed to have periodic boundary conditions, namely $\hat{\sigma}_{N+1}^{\alpha}$
should be identified with $\hat{\sigma}_{1}^{\alpha}$.

\subsection{The variational principle\label{subsec:var_princ}}

We begin by formulating the conventional variational principle in
terms of the density matrix. The density matrix for the system can
most generally be written as

\begin{equation}
\hat{\rho}=\frac{1}{2^{N}}\sum_{\alpha_{1},\dots,\alpha_{N}}p_{\alpha_{1},\dots,\alpha_{N}}\hat{\sigma}_{1}^{\alpha_{1}}\hat{\sigma}_{2}^{\alpha_{2}}\dots\hat{\sigma}_{N}^{\alpha_{N}},
\end{equation}
where each of the indices $\alpha_{n}$ runs over $0,x,y,z$. In this
representation the density matrix is parameterized by the tensor $p_{\alpha_{1},\dots,\alpha_{N}}$,
which contains $4^{N}$ elements. For $\hat{\rho}$ to be a valid
density matrix it must obey $\rho=\rho^{\dagger},$ ${\rm Tr}(\rho)=1$,
and $\rho\succeq0$ (positive semidefiniteness). The first two conditions
are enforced by $p_{\alpha_{1},\dots,\alpha_{N}}=p_{\alpha_{1},\dots,\alpha_{N}}^{\ast}$
and $p_{0,\dots,0}=1$, respectively.

The variational principle states that the ground state of $\hat{H}$
is described by the parameters, $p_{\alpha_{1},\dots,\alpha_{N}}^{{\rm gs}}$,
satisfying
\begin{equation}
\begin{split}\underset{\{p\}}{{\rm minimize}}\,\, & E(\{p\})={\rm Tr}(\hat{\rho}\hat{H})\\
{\rm s.t.}\hspace{1em} & \hspace{1em}\hat{\rho}(\{p\})\succeq0.
\end{split}
\label{eq:var_princ}
\end{equation}
In the absence of ground-state degeneracy, the resulting density matrix
is guaranteed to describe a pure state, while in the case of a degenerate
ground state it can more generally describe a classical mixture of
several ground states.

For a large system, obtaining $p_{\alpha_{1},\dots,\alpha_{N}}^{{\rm gs}}$
through a straight-forward numerical minimization is impractical due
to the exponentially-large number of parameters in $\{p\}$. Notice,
however, that the function to be minimized, $E(\{p\})$, only involves
a small subset of the elements in $\{p\}$. Specifically, it contains
those elements that have the form 
\begin{equation}
{\rm Tr}(\hat{\rho}\hat{\sigma}_{n}^{i}\hat{\sigma}_{n+1}^{\alpha})=p_{\underset{1}{0},0,\dots,0,\underset{n}{i},\alpha,0,\dots,\underset{N}{0}},
\end{equation}
whose number scales linearly with the system size, $N$. Instead,
it is the condition $\hat{\rho}\succeq0$ that limits the application
of the variational principle by coupling this subset with the rest
of the parameters in $\{p\}$.

With this in mind, we now look for a condition that would approximate
$\hat{\rho}\succeq0$ in a manner involving only a subset of parameters
in $\{p\}$. More specifically, we shall formulate an ordered approximation
to $\hat{\rho}\succeq0$ that invokes a number of parameters scaling
quadratically in $N$, for any order of the approximation (and linearly
in $N$ for translationally-invariant systems).

\subsection{Approximating the positive-semidefiniteness condition\label{subsec:approx_PSD}}

To approximate the condition $\hat{\rho}\succeq0$, we rely on the
observation that $\hat{\rho}\succeq0$ \emph{if and only if} 

\emph{
\begin{align}
{\rm Var}(\hat{O}) & =\langle(\hat{O}-\langle\hat{O}\rangle)^{2}\rangle\ge0\label{eq:rho_psd_equiv_Var_o}
\end{align}
}for any hermitian operator $\hat{O}$, where $\left\langle \cdot\right\rangle \equiv{\rm Tr}(\hat{\rho}\,\cdot)$.
The forward direction easily follows from noting that ${\rm Var}(\hat{O})={\rm Tr}(\hat{\rho}\hat{A})$
where $\hat{A}=[\hat{O}-{\rm Tr}(\hat{\rho}\hat{O})]^{2}$ is positive
semidefinite, because it is the square of a Hermitian operator. Together
with $\hat{\rho}\succeq0$ this readily ensures ${\rm Var}(\hat{O})\ge0$.

To prove the converse direction, let us denote by $\{|l\rangle,w_{l}\}_{l}$
the set of eigenstates and eigenvalues of $\hat{\rho}$. If ${\rm Var}(\hat{O})\ge0$
for any hermitian $\hat{O}$, it is in particular true for $\hat{O}=\left|l\right\rangle \left\langle l\right|$,
from which it follows that
\begin{equation}
0\le{\rm Var}(\left|l\right\rangle \left\langle l\right|)=w_{l}(1-w_{l}),
\end{equation}
namely $0\le w_{l}\le1$. Since this is true for any $l$, one concludes
that all the eigenvalues of $\hat{\rho}$ are non-negative, i.e. $\hat{\rho}\succeq0$.

The equivalence between $\hat{\rho}\succeq0$ and Eq.~(\ref{eq:rho_psd_equiv_Var_o})
suggests a route towards approximating the condition $\hat{\rho}\succeq0$.
Instead of requiring that all hermitian operators have a non-negative
variance, let us limit this requirement to the subset of hermitian
operators that have the form

\begin{equation}
\hat{O}^{(k)}=\sum_{n,\nu}c_{n\nu}\hat{L}_{n,\nu}^{(k)},\label{eq:O_k}
\end{equation}
where $c_{n\nu}$ are real coefficients and $\{\hat{L}_{n,\nu}^{(k)}\}_{\nu=1}^{3\times4^{k-1}}$
span the space of range-$k$ local hermitian zero-trace operators,
that is
\begin{equation}
\begin{split}\{\hat{L}_{n,\nu}^{(1)}\}_{\nu} & =\{\hat{\sigma}_{n}^{i}\}_{i}\,\,\,;\,\,\,\{\hat{L}_{n,\nu}^{(2)}\}_{\nu}=\{\hat{\sigma}_{n}^{i}\hat{\sigma}_{n+1}^{\alpha}\}_{i,\alpha}\,\,\,;\\
\{\hat{L}_{n,\nu}^{(3)}\}_{\nu} & =\{\hat{\sigma}_{n}^{i}\hat{\sigma}_{n+1}^{\alpha}\hat{\sigma}_{n+2}^{\beta}\}_{i,\alpha,\beta}\,\,\,;\,\,\,\dots
\end{split}
\label{eq:Lk_def}
\end{equation}
where $i\in\{x,y,z\}$ and $\alpha,\beta\in\{0,x,y,z\}$. As $k$
increases, the condition ${\rm Var}(\hat{O}^{(k)})\ge0$ becomes a
better approximation of $\hat{\rho}\succeq0$. While the two conditions
are strictly equivalent only when $k=N$, we shall see below that
it is often sufficient to consider the case $k=2$ or $k=3$. 

The benefit of using the condition ${\rm Var}(\hat{O}^{(k)})\ge0$
instead of the exact condition, $\hat{\rho}\succeq0$, is that it
can be enforced by constraining a relatively small set of the parameters,
whose number scales only quadratically with the system size, $N$.
To see this, let us substitute the expression for $\hat{O}^{(k)}$,
Eq.~(\ref{eq:O_k}), in the condition ${\rm Var}(\hat{O}^{(k)})\ge0$.
This results in the condition
\begin{equation}
\sum_{m,n,\mu,\nu}c_{m\mu}\mathcal{M}_{m\mu,n\nu}^{(k)}c_{n\nu}\ge0\hspace{1em}\forall\,\,\{c_{n\nu}\},\label{eq:cMcg0}
\end{equation}
where

\begin{equation}
\begin{split}\mathcal{M}_{m\mu,n\nu}^{(k)}\equiv\frac{1}{2} & \langle\{\hat{L}_{m\mu}^{(k)},\hat{L}_{n\nu}^{(k)}\}\rangle-\langle\hat{L}_{m\mu}^{(k)}\rangle\langle\hat{L}_{n\nu}^{(k)}\rangle,\end{split}
\label{eq:M_k}
\end{equation}
is the correlation matrix considered in Ref.~\citep{Qi2019Determining},
which is manifestly real and symmetric. Here, $\{\cdot,\cdot\}$ stands
for the anticommutator. The condition, Eq.~(\ref{eq:cMcg0}), is
equivalent to $\mathcal{M}^{(k)}\succeq0$. We thus approximate the
condition that $\hat{\rho}$ is positive semidefinite by the condition
that the correlation matrix, $\mathcal{M}^{(k)}$, is positive semidefinite.
Importantly, while the dimension of $\hat{\rho}$ is $d_{\rho}=2^{N}$,
making it intractable, the dimension of $\mathcal{M}^{(k)}$ is $d_{\mathcal{M}}=3\cdot4^{k-1}N$.

Had we not limited the range of the operators $\hat{L}^{(k)}$ in
Eq.~(\ref{eq:O_k}), the resulting constraint would be equivalent
to the so-called $k$-positivity conditions~\citep{Mazziotti2001Uncertainty},
which require that all $k$-body reduced density matrices (RDM) are
positive-semidefinite, and which are at the source of the variational
2-RDM method~\citep{Mazziotti2002variational,Zhao2004reduced,Mazziotti2005variational,Mazziotti2006quantum,Mazziotti2007reduced,Barthel2012Solving,Baumgratz2012Lower,Anderson2013second,Verstichel2013Extensive,Mazziotti2016enhanced,Alcoba2018direct,Rubio2019variational}.
The dimension of the $k$-body RDM scales as $N^{k}$, which typically
limits the ability to increase accuracy by increasing $k$. In contrast,
by restricting the range of $\hat{L}^{(k)}$ to be $k$, we force
the dimension of the correlation matrix $\mathcal{M}^{(k)}$ to remain
linear in $N$ for any fixed $k$. In this regard, one should also
note Ref.~\citep{Barthel2012Solving}, where restrictions on the
range of considered operators are placed using a different protocol,
enabling the authors to treat both 1d and 2d lattice systems. Finally,
note that although the range $\hat{L}^{(k)}$ is restricted to $k$,
the matrix $\mathcal{M}^{(k)}$ contains also long-range correlations
since $m$ and $n$ in Eq.~(\ref{eq:M_k}) are not restricted.

\subsection{The variational-correlation procedure\label{subsec:The_VC_methos}}

We are now in a position to describe the variational-correlation procedure
for approximating the system's ground-state. We define the variational
parameters as the disconnected correlation functions

\begin{align}
b_{n\nu}\equiv\langle\hat{L}_{n\nu}^{(k)}\rangle\,\,\,\,;\,\,\,\,C_{m\mu n\nu} & \equiv\langle\hat{L}_{m\mu}^{(k)}\hat{L}_{n\nu}^{(k)}\rangle\,\:{\rm for}\,\,(n\ge m+k),\label{eq:B_C_def}
\end{align}
collectively denoted by $x=\{b_{n\nu},C_{\mu\nu}^{mn}\}$, whose number
scales quadratically with $N$. The approximate ground-state energy
and correlations are then obtained ``to order $k$'' by solving the
following optimization problem:
\begin{equation}
\begin{split} & \underset{\{x\}}{{\rm minimize}}\,\,\,E(\{x\})={\rm Tr}(\hat{\rho}\hat{H}),\\
 & {\rm s.t.}\hspace{1em}\mathcal{M}^{(k)}(\{x\})\succeq0.
\end{split}
\label{eq:VC_opt_prob}
\end{equation}

The resulting energy, $E(\{x_{{\rm min}}\})$, sets a \emph{lower
bound} on the true ground-state energy of the system. This is because
the constraint in Eq.~(\ref{eq:VC_opt_prob}) is a result of relaxing
the constraint in the original minimization problem of Eq.~(\ref{eq:var_princ}).
Namely, the minimum point of Eq.~(\ref{eq:var_princ}), which describes
the exact ground state, is \emph{contained} within the space of feasible
points considered in the minimization problem of Eq.~(\ref{eq:VC_opt_prob}).
Alternatively stated, unlike the conventional use of the variational
principle where one places additional constrains on the wave function, here one relaxes the constraints on it (by replacing $\hat{\rho}\succeq0$
with $\mathcal{M}^{(k)}\succeq0$). While the former procedure yields
an energy which is greater than the ground state, the latter yields
an energy which is lower.

This, of course, comes at a price. First, we do not possess all the
information about the ground state, but rather only correlation functions
of the form given in Eq.~(\ref{eq:B_C_def}). Second, the resulting
ground state is generally not physical. In other words, the correlation
functions $x_{{\rm min}}=\{b,C\}_{{\rm min}}$ cannot arise from an
exactly positive semidefinite $\hat{\rho}$. Nevertheless, as $k$
increases these correlations should approximate the true ground-state
correlation functions with increasing accuracy. In Sec.~\ref{sec:Num_Results},
we demonstrate this method up to order $k=3$.

\subsection{Interpretation of active constraints\label{subsec:active_constr}}

Since the objective function in Eq.~(\ref{eq:VC_opt_prob}) is linear,
the minimum point is always at the boundary of the region defined
by $\mathcal{M}^{(k)}\succeq0$. Namely, at the optimum, at least
some of the eigenvalues of $\mathcal{M}^{(k)}$ are zero; these represent
the \emph{active} constraints of the problem. Each such zero eigenvalue
of $\mathcal{M}^{(k)}$ corresponds to an eigenvector, with elements
$\omega_{n\nu}$, which can be used to define an operator $\hat{\Omega}\in\hat{O}^{(k)}$,
\begin{equation}
\hat{\Omega}=\sum_{n,\nu}\omega_{n\nu}\hat{L}_{n\nu}^{(k)}.
\end{equation}
Each such operator then obeys
\begin{equation}
\langle\hat{\Omega}^{2}\rangle-\langle\hat{\Omega}\rangle^{2}=\sum_{m,n,\mu,\nu}\omega_{m\mu}\mathcal{M}_{m\mu,n\nu}^{(k)}\omega_{n\nu}=0,\label{eq:ActiveConstraints}
\end{equation}
due to $\omega_{mn}$ being in the null space of $\mathcal{M}_{m\mu,n\nu}^{(k)}$.
From Eq.~(\ref{eq:ActiveConstraints}), one infers that the ground
state is an eigenstate of $\hat{\Omega}$. This could suggest that
the accuracy of the variational correlation approximation is determined
by the locality of the operators for which the ground-state is an
eigenstate. If these operators can be approximated by operators in
$\hat{O}^{(k)}$, they would manifest as active constraints in the
minimization of Eq.~(\ref{eq:VC_opt_prob}), forcing the energy to
increase and thereby get closer to the exact ground state energy.

\subsection{Translational Invariance\label{subsec:trans_inv}}

The variational procedure described above can be significantly simplified
when the Hamiltonian of Eq.~(\ref{eq:Hamiltonian}) is translationally-invariant,
namely when $J_{ni\alpha}$ is independent of $n$. In this case,
the ground state (and in fact any eigenstate) of $\hat{H}$ is guaranteed
to either be translationally invariant or degenerate. In the latter
case one can always choose a superposition (or a classical mixture)
of the degenerate ground states that would itself be translationally
invariant.

We can, therefore, impose translational invariance on the variational
correlations,

\begin{align}
C_{m\mu;n\nu} & =C_{\mu\nu}^{\left(n-m\right)}\hspace{1em};\hspace{1em}b_{n\nu}=b_{\nu}.
\end{align}
This reduces the number of variational parameters, now denoted by
$x=\{b_{\nu},C_{\mu\nu}^{\Delta n}\}$. In particular, the number
of elements in $x$ scales only linearly with $N$. 

\section{Numerical Results\label{sec:Num_Results}}

In this section, we demonstrate the application of the variational
correlation (VC) approximation, as defined in Eq.~(\ref{eq:VC_opt_prob}),
for studying 1d Hamiltonians of nearest-neighbor interacting spin-$1/2$
particles. We focus on two types of models: (i) the tilted-field Ising
model and (ii) the XXZ model. 

We begin by examining the results for the ground state energy within
the $k=2$ order of the approximation. These results are compared
with those obtained from the \emph{density matrix renormalization
group} (DMRG), which for the studied systems are essentially exact.
Next, the error within the $k=2$ order is examined as a function
of the systems size and upon introducing disorder. We then study the
dependence of the ground-state energy on $k$ by comparing results
for $k=1,2,3$. Finally, we examine the results for the correlation
functions. Details regarding the numerical implementation of the VC
approximation are found in Appendix~\ref{sec:Implementation}.

\subsection{Ground-state energy\label{subsec:Num_res_E_gs}}

\subsubsection{Tilted-field Ising model}

The Ising model with a tilted-filed is described by the Hamiltonian
\begin{equation}
\hat{H}_{{\rm Ising}}=\sum_{n=1}^{N}\left(h_{x}\hat{\sigma}_{n}^{x}+h_{z}\hat{\sigma}_{n}^{z}+J_{z}\hat{\sigma}_{n}^{z}\hat{\sigma}_{n+1}^{z}\right),\label{eq:H_tilted_field_Ising}
\end{equation}
where periodic boundary conditions are assumed. This Hamiltonian is
a special case of the Hamiltonian considered in Eq.~(\ref{eq:Hamiltonian}).
The direction of the field in the $xz$ plane is parameterized by
the angle, $\theta,$ where $h_{x}=|\vec{h}|\cos\theta$ and $h_{z}=|\vec{h}|\sin\theta$.

\begin{figure}
\begin{centering}
\begin{tabular}{lr}
\hskip -2mm
\includegraphics[clip=true,trim=0mm 0mm 0mm 0mm,height=3.4cm]{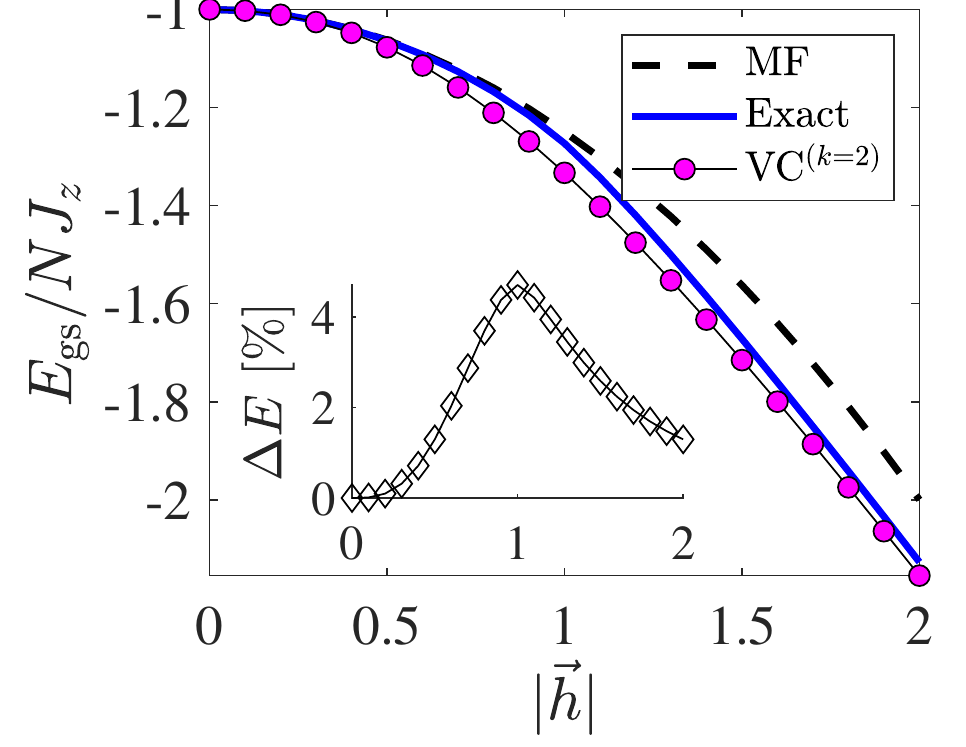}
\llap{\hskip 2mm \parbox[c]{8.5cm}{\vspace{-6.2cm}(a)}}
&
\hskip -1mm
\includegraphics[clip=true,trim=0mm 0mm 0mm 0mm,height=3.4cm]{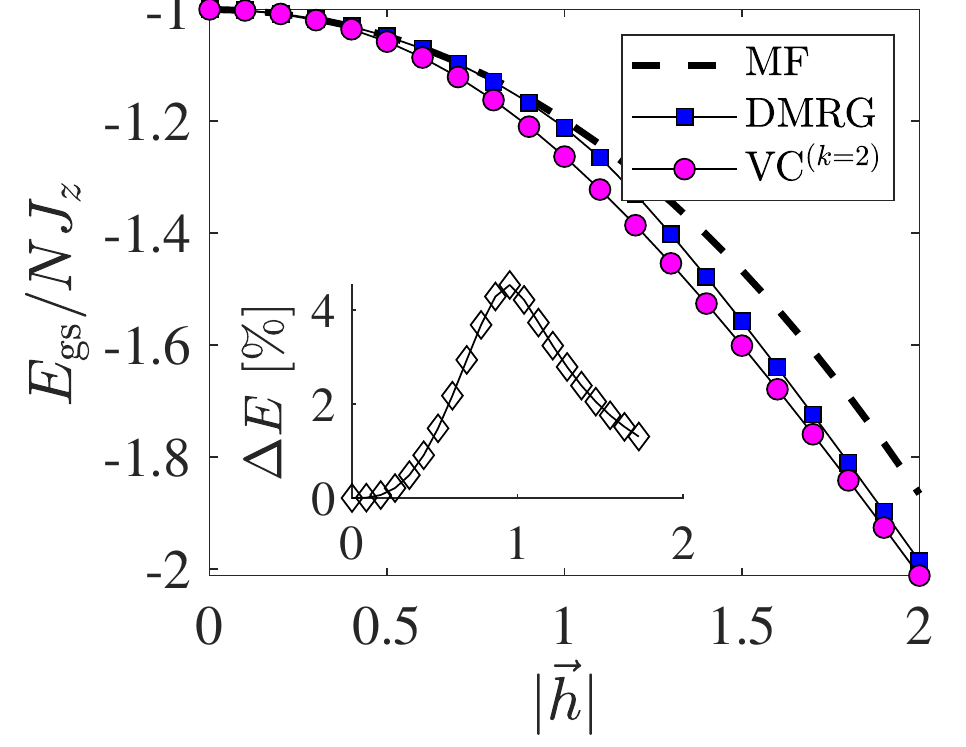}
\llap{\hskip 2mm \parbox[c]{8.5cm}{\vspace{-6.2cm}(b)}}
\end{tabular}
\end{centering}

\caption{Ground-state energy of the Ising model {[}see Eq.~(\ref{eq:H_tilted_field_Ising}){]}
within the variational-correlations (VC) approximation, to order $k=2$,
as a function of the field $|\vec{h}|$. The system size is $N=80$
spins. The VC method sets a lower-bound on the exact ground state
energy of the system. In (a), the field angle is $\theta=0$, allowing
for an exact analytical solution (solid blue line) by virtue of the
Jordan Wigner transformation. In (b), the field angle is $\theta=\pi/6$
and results are compared with a density matrix renormalization group
(DMRG) calculation (blue squares). Results are also compared with
a mean-filed calculation (dashed black line), which sets an upper
bound on the ground state energy.\label{fig:Ising_E}}
\end{figure}
In Figs.~\hyperref[fig:Ising_E]{\ref{fig:Ising_E}(a)} and ~\hyperref[fig:Ising_E]{\ref{fig:Ising_E}(b)},
we present the ground-state energy for $\theta=0$ and $\theta=\pi/6$,
respectively, calculated using the VC approximation to order $k=2$,
for a system of $N=80$ spins. The results are shown as a function
of the field strength, $|\vec{h}|$, for fixed $J_{z}=1$. In the
case of $\theta=0$, the Hamiltonian can be solved exactly by mapping
the problem to a system of free fermions by virtue of the Jordan-Wigner
transformation~\citep{Suzuki2012quantum}. This solution is marked
by a solid blue line in Fig.~\hyperref[fig:Ising_E]{\ref{fig:Ising_E}(a)}.
As can be seen, the VC approximation is in reasonable agreement with
the exact solution already in the $k=2$ order. Notice the maximal
discrepancy is at $|\vec{h}|=J_{z},$ where the system is known to
go through a continuous phase transition. 

For a general field angle, an exact analytical calculation is not
possible. Accordingly, the results of the VC approximation for the
case of $\theta=\pi/6$, shown in Fig.~\hyperref[fig:Ising_E]{\ref{fig:Ising_E}(b)},
are compared with those of a DMRG calculation. The latter is implemented
using the iTensor library~\citep{ITensor}. The results are qualitatively
similar to those obtained for $\theta=0$. Importantly, we see that
deviating from integrability does not reduce the accuracy of the approximation. 

As explained in Sec.~\ref{subsec:The_VC_methos}, the variational
correlation approximation sets a lower bound on the ground-state energy,
contrary to the conventional variational principle which allows one
to obtain an upper bound. This is manifested in comparing the results
with those of a mean-field calculation. The latter is obtained by
considering a product-state trial variational wave function, $|\Psi_{{\rm MF}}\rangle$,
and minimizing $\langle\Psi_{{\rm MF}}|\hat{H}|\Psi_{{\rm MF}}\rangle$.
Indeed, the mean-field result (dashed black line) bounds the exact
result from above, while the VC approximation bounds it from below.

\subsubsection{The XXZ model}

\begin{figure}
\begin{centering}
\begin{tabular}{lr}
\hskip -3mm
\includegraphics[clip=true,trim=0mm 0mm 0mm 0mm,height=3.4cm]{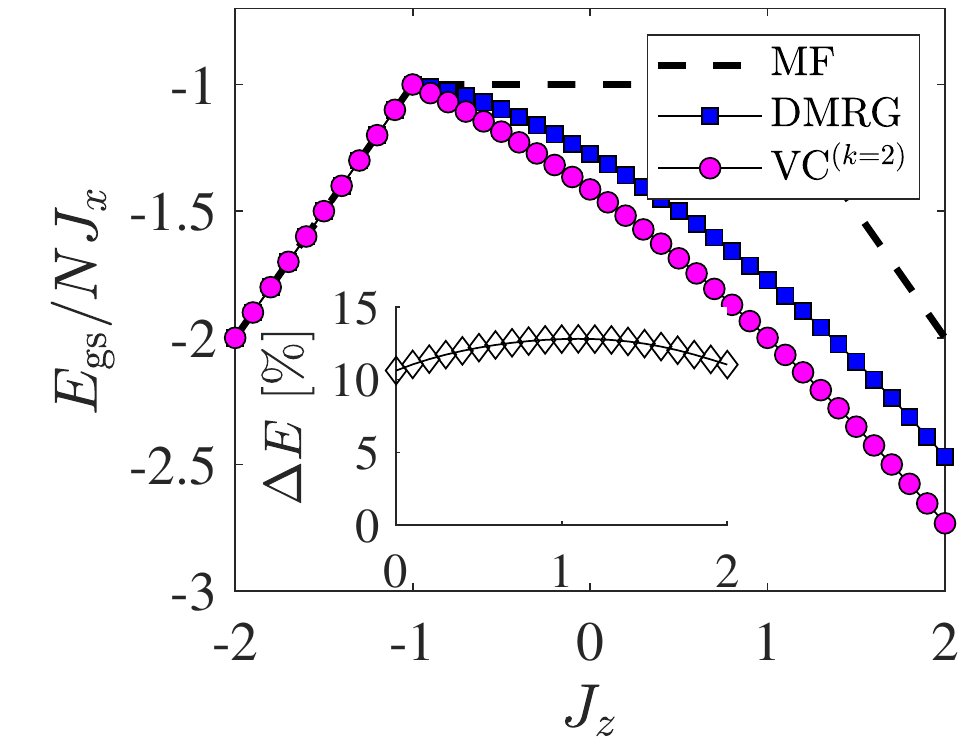}
\llap{\hskip 2mm \parbox[c]{8.1cm}{\vspace{-0.1cm}(a)}}
&
\hskip -2mm
\includegraphics[clip=true,trim=0mm 0mm 0mm 0mm,height=3.4cm]{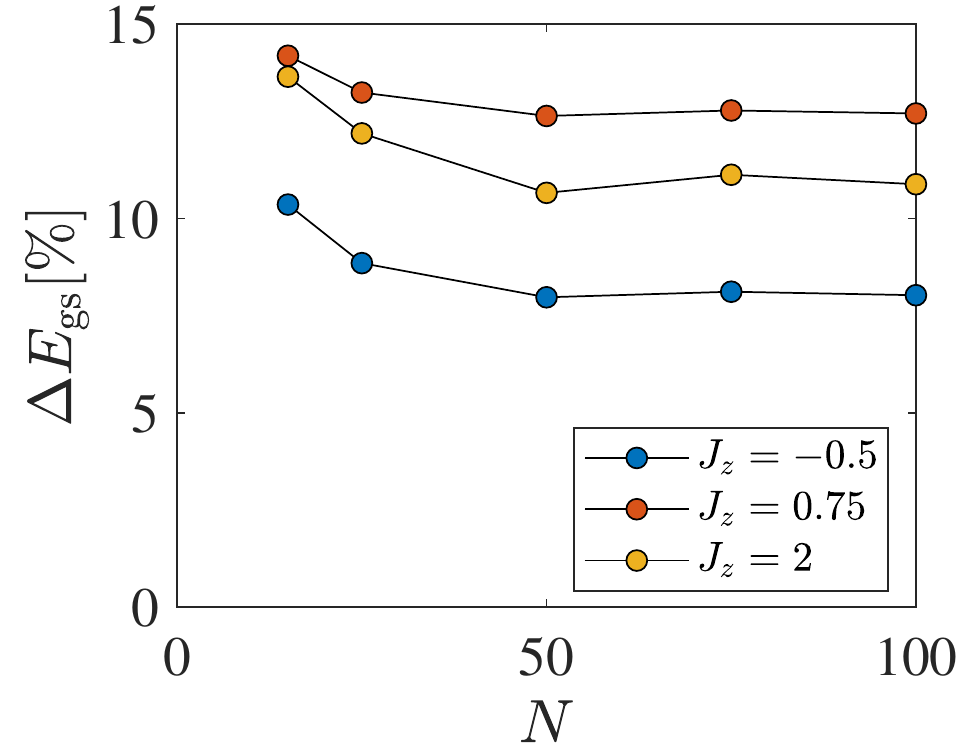}
\llap{\hskip 2mm \parbox[c]{8.2cm}{\vspace{-0.1cm}(b)}}
\end{tabular}
\end{centering}

\caption{(a) Ground-state energy of the XXZ model with $N=80$ spins {[}see
Eq.~(\ref{eq:H_XXZ}){]}, within the variational-correlations (VC)
approximation, calculated to order $k=2$. Results are compared with
a DMRG calculation (blue squares) and with a mean-filed calculation
(dashed black line). At $J_{z}=-1$, the system goes through a first-order
quantum phase transition, above which a discrepancy of up to $\sim12\%$
develops between the VC and the DMRG results. (b) Difference between
the ground-state energy calculated by DMRG and by the VC method to
order $k=2$, as a function of system size, $N$, for fixed $J_{x}=1$
and different values of $J_{z}$. Notice the error does not increase
with system size. As shown below, in Fig.~\ref{fig:E_gs_vs_k}, this
error of the VC approximation decreases significantly upon going to
the $k=3$ order.\label{fig:XXZ_E}}
\end{figure}

The Hamiltonian describing the XXZ model is given by
\begin{equation}
\hat{H}_{{\rm XXZ}}=\sum_{n=1}^{N}\left[J_{x}\left(\hat{\sigma}_{n}^{x}\hat{\sigma}_{n+1}^{x}+\hat{\sigma}_{n}^{y}\hat{\sigma}_{n+1}^{y}\right)+J_{z}\hat{\sigma}_{n}^{z}\hat{\sigma}_{n+1}^{z}\right],\label{eq:H_XXZ}
\end{equation}
where, as before, periodic boundary conditions are assumed. The results
for the ground-state energy as a function of the field $J_{z}$ are
shown in Fig.~\hyperref[fig:XXZ_E]{\ref{fig:XXZ_E}(a)} for fixed
$J_{x}=1$ and $N=80$. As before, we compare the variational correlation
method, calculated to order $k=2$, with the result of the variational
mean-field state (dashed black line) and of the DMRG calculation (blue
squares).

For $J_{z}<-1$, the ground state is a symmetry-broken ferromagnetic
state with all spins pointing in the $z$ direction, and the energy
is, therefore, linear in $J_{z}$. In this phase the VC, DMRG, and
mean-field calculations all coincide. At $J_{z}=-1$, a first-order
quantum phase transition occurs signalled by the discontinuous derivative
of the ground-state energy. For $J_{z}>-1$, the results of the VC
calculation start to deviate considerably from the DMRG result. In
Sec.~\ref{subsec:Num_res_k_depnd} we shall see that this discrepancy
can be mitigated by increasing the approximation order to $k=3$.

\subsection{Scaling of the error with system size\label{subsec:finite_size_scale}}

Before moving on to study the effect of increasing the order of approximation,
$k$, it is important to examine the error of the VC approximation
as a function of the system size, for a fixed value of $k$.

In Fig.~\hyperref[fig:XXZ_E]{\ref{fig:XXZ_E}(b)}, we present the
relative energy difference $\Delta E=(E_{{\rm dmrg}}-E_{{\rm vc}})/|E_{{\rm dmrg}}|$
between the DMRG result and the VC approximation, calculated to order
$k=2$, as a function of the number of spins, $N$, for the XXZ model
of Eq.~(\ref{eq:H_XXZ}). Results are shown for several different
values of $J_{z}$. Importantly, the relative error in energy does
not increase with system size but rather goes to a constant.

\subsection{Disorder}

Next, let us examine the effect of introducing a disordered field.
This is done by adding a term
\begin{equation}
\hat{H}_{{\rm dis}}=\sum_{n=1}^{N}h_{n}^{{\rm dis}}\hat{\sigma}_{n}^{z},\label{eq:H_dis}
\end{equation}
to the Hamiltonians of Eqs.~(\ref{eq:H_tilted_field_Ising}) and~(\ref{eq:H_XXZ}),
where $h_{n}^{{\rm dis}}\in[-W_{{\rm dis}},W_{{\rm dis}}]$ is a uniformly
distributed random variable and $N=80$.

\begin{figure}
\begin{centering}
\begin{tabular}{lr}
\hskip -2mm
\includegraphics[clip=true,trim=0mm 0mm 0mm 0mm,height=3.4cm]{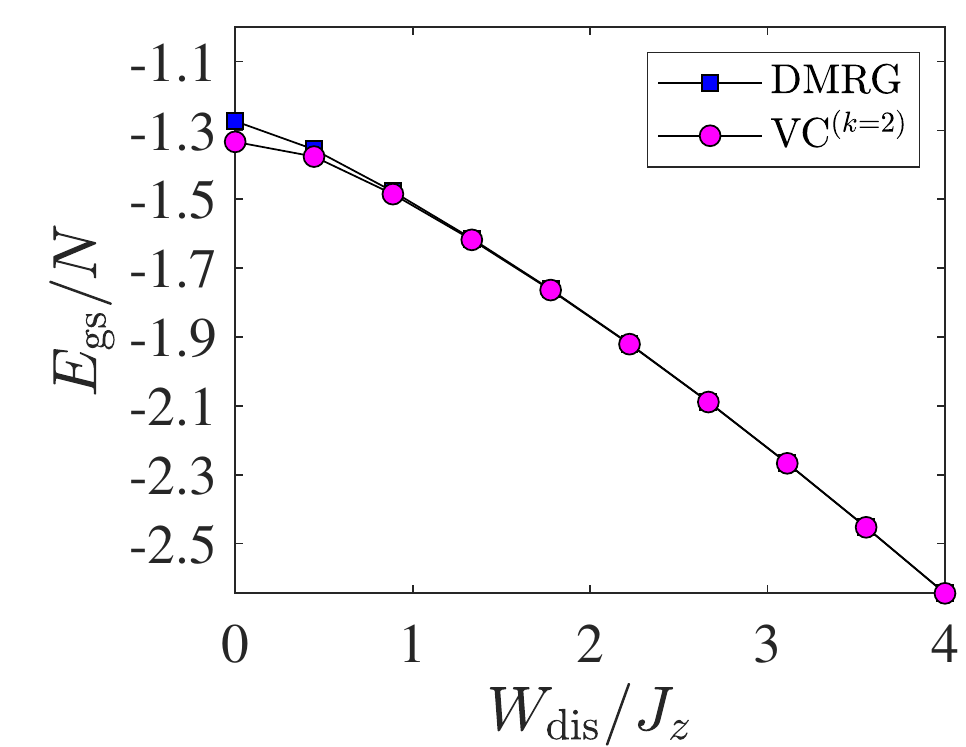}
\llap{\hskip 2mm \parbox[c]{8.5cm}{\vspace{-6.2cm}(a)}}
&
\hskip -2mm
\includegraphics[clip=true,trim=0mm 0mm 0mm 0mm,height=3.4cm]{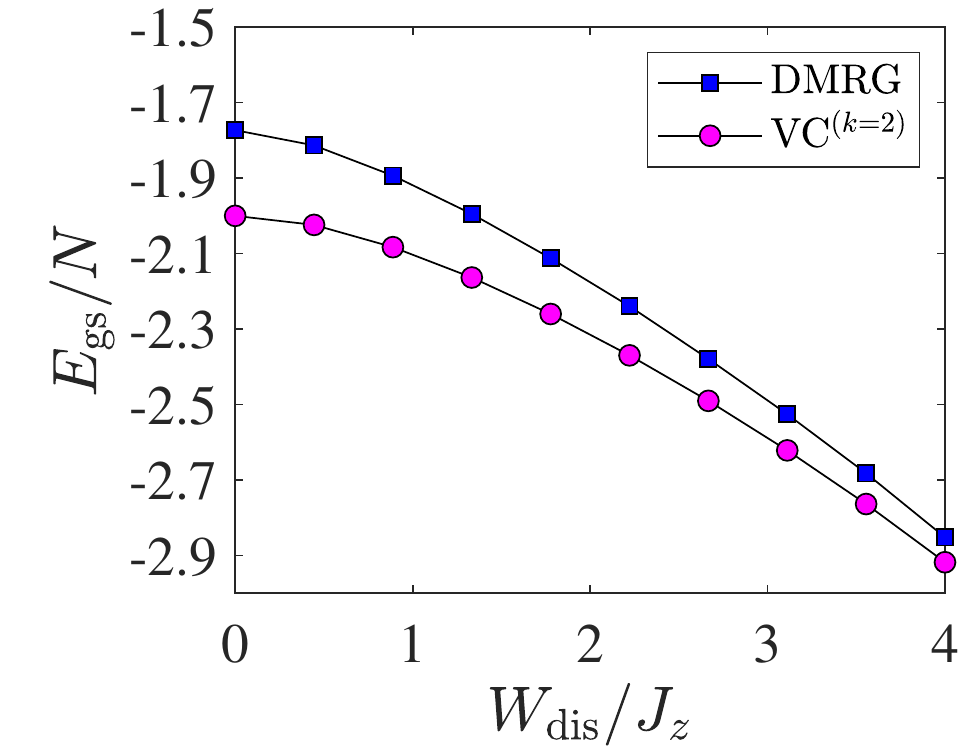}
\llap{\hskip 2mm \parbox[c]{8.5cm}{\vspace{-6.2cm}(b)}}
\end{tabular}
\end{centering}

\caption{Effect of a disorder field in the $z$ direction. The ground-state
energy is plotted versus the disorder strength, as defined below Eq.~(\ref{eq:H_dis}),
for (a) the transverse-field Ising model at its critical point, $h_{x}=J_{z}$,
and (b) the Heisenberg model, $J_{x}=J_{z}$. The plots correspond
to a single disorder realization. For both models, the variational
correlation (VC) approximation becomes better as the disorder strength
increases. Here, the system size was taken to be $N=80$. \label{fig:E_gs_vs_disorder}}
\end{figure}

In Fig.~\ref{fig:E_gs_vs_disorder}, the ground-state energy is presented
as a function of the disorder strength, $W_{{\rm dis}}$, for a single
disorder realization, calculated using both DMRG and the VC approximation
to order $k=2$. Figure~\hyperref[fig:E_gs_vs_disorder]{\ref{fig:E_gs_vs_disorder}(a)}
shows results for the transverse-field Ising model at its critical
point, $h_{x}=J_{z}$, and Fig.~\hyperref[fig:E_gs_vs_disorder]{\ref{fig:E_gs_vs_disorder}(b)}
shows results for the Heisenberg model, obtained by setting $J_{x}=J_{z}$
in Eq.~(\ref{eq:H_XXZ}).

Increasing the strength of disorder actually improves the performance
of the VC method at predicting true ground-state energies (i.e. the
DMRG result) in both models. This could be related to the localization
induced by the disorder field (see also the discussion in Sec.~\ref{subsec:active_constr}).

\subsection{Dependence on approximation order\label{subsec:Num_res_k_depnd}}

\begin{figure}
\begin{centering}
\begin{tabular}{lr}
\hskip -2mm
\includegraphics[clip=true,trim=0mm 0mm 0mm 0mm,height=3.4cm]{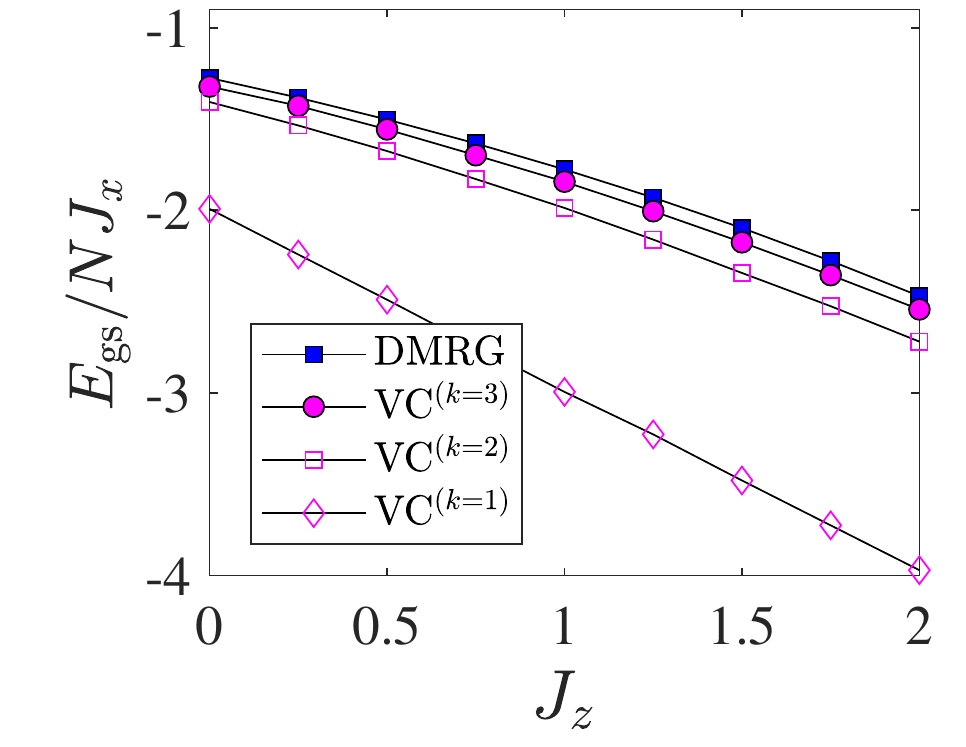}
\llap{\hskip 2mm \parbox[c]{8.5cm}{\vspace{-6.2cm}(a)}}
&
\hskip 3mm
\includegraphics[clip=true,trim=10mm 0mm 0mm 2mm,height=3.32cm]{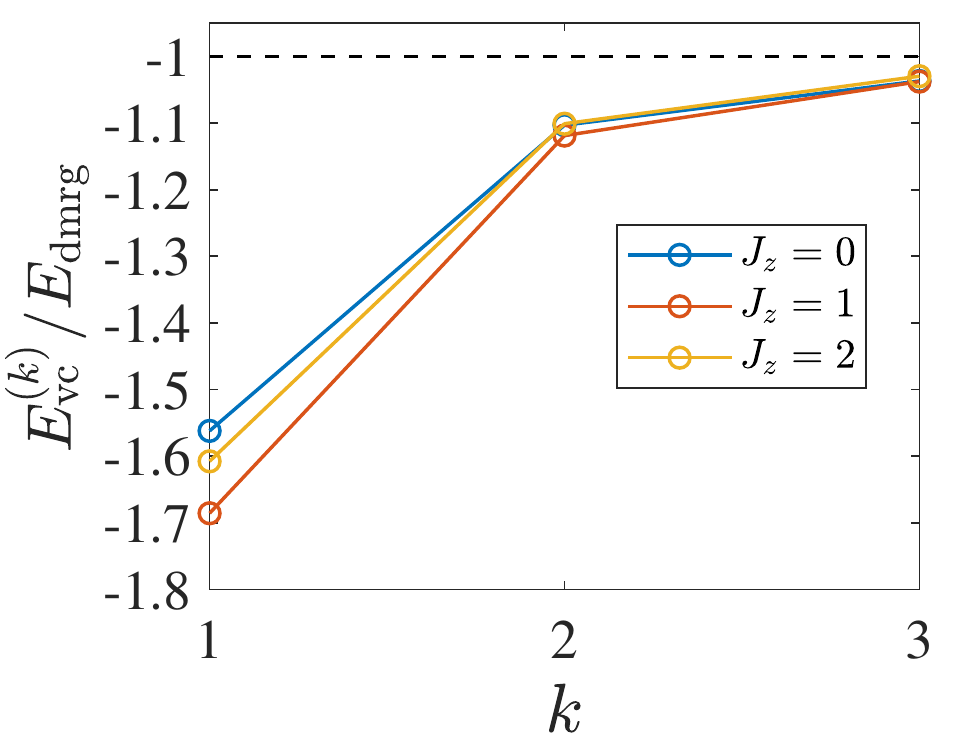}
\llap{\hskip 2mm \parbox[c]{8.5cm}{\vspace{-6.2cm}(b)}}
\llap{\parbox[c]{8.85cm}{\vspace{-3.5cm}\rotatebox{90}{$E_{\rm vc}^{(k)}/E_{\rm dmrg}$}}}
\\
\hskip -2mm
\includegraphics[clip=true,trim=0mm 0mm 0mm 0mm,height=3.4cm]{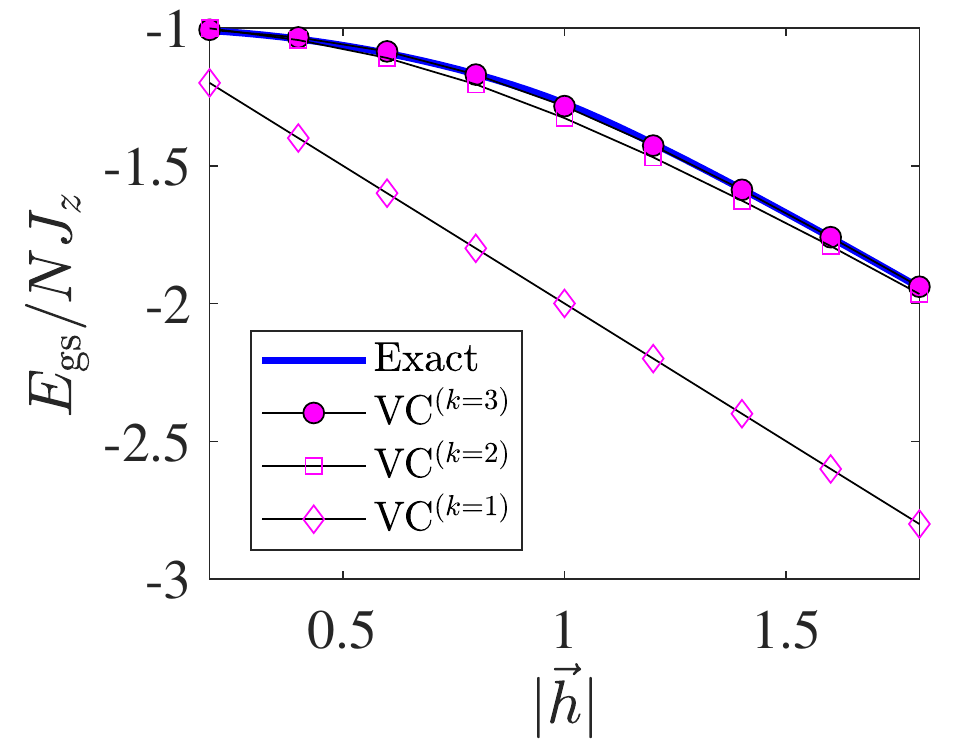}
\llap{\hskip 2mm \parbox[c]{8.5cm}{\vspace{-6.2cm}(c)}}
&
\hskip 3mm
\includegraphics[clip=true,trim=10mm 0mm 0mm 2mm,height=3.32cm]{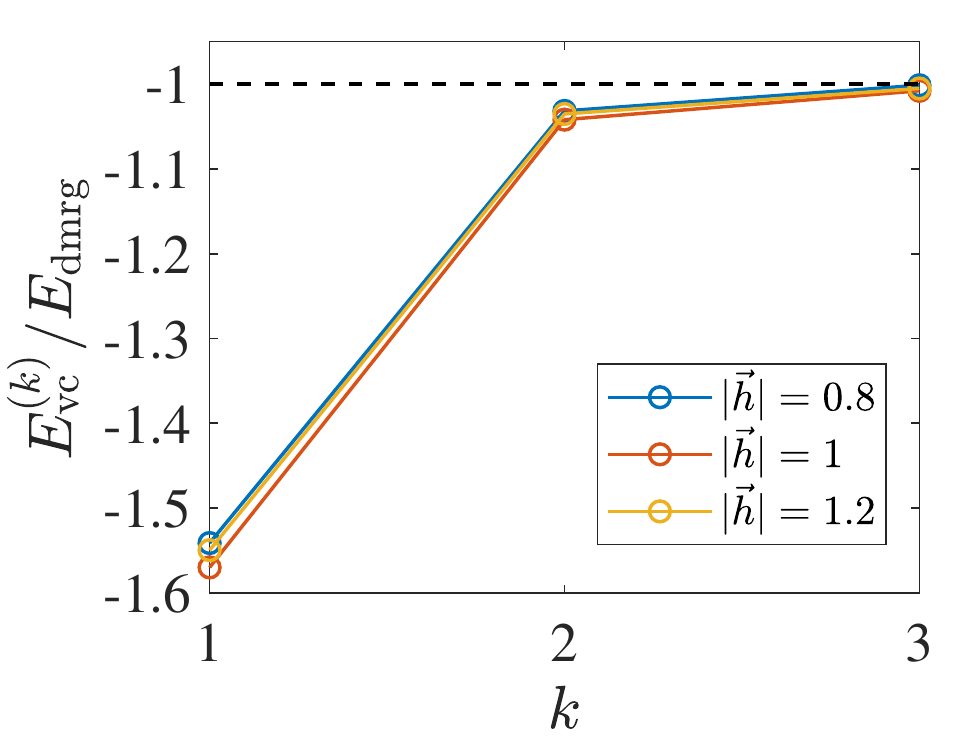}
\llap{\hskip 2mm \parbox[c]{8.5cm}{\vspace{-6.2cm}(d)}}
\llap{\parbox[c]{8.85cm}{\vspace{-3.5cm}\rotatebox{90}{$E_{\rm vc}^{(k)}/E_{\rm exact}$}}}
\end{tabular}
\end{centering}

\caption{(a) Ground-state energy within the variational correlation (VC) approximation
for the XXZ model with $N=30$ spins, for different values of the
approximation order, $k=1,2,3$, with fixed $J_{x}=1$. Upon increasing
$k$, the VC results approach those of the DMRG calculation, shown
in blue squares. In (b), the energy is plotted as a function of $k$,
for several values of $J_{z}$. Similar behavior is observed for the
the transverse-field Ising model in (c) and (d).\label{fig:E_gs_vs_k}}
\end{figure}

We now study the dependence of the VC approximation on the order of
approximation, $k$. We examine first the XXZ model, for which the
$k=2$ results were presented in Fig.~\ref{fig:XXZ_E}. In Fig.~\hyperref[fig:E_gs_vs_k]{\ref{fig:E_gs_vs_k}(a)},
we present the results for the ground state energy with $N=30$ spins,
calculated within the VC approximation to orders $k=1,2,3$. As $k$
increases, the energy approaches the DMRG results, shown in blue squares.
This is emphasized in Fig.~\hyperref[fig:E_gs_vs_k]{\ref{fig:E_gs_vs_k}(b)}
which presents the VC ground state energy, $E_{{\rm vc}}$, normalized
by $E_{{\rm dmrg}}$, for three values of $J_{z}$ as a function of
$k$.

Similarly, In Figs.~\hyperref[fig:E_gs_vs_k]{\ref{fig:E_gs_vs_k}(c,d)}
the ground-state energy for the transverse-field Ising model ($\theta=0$)
with $N=30$ spins is examined for three different approximation orders,
$k=1,2,3$. Qualitatively similar behavior as in the XXZ model is
observed, although with a faster convergence.

\subsection{Correlation functions\label{subsec:Corr_func}}

The VC approach, as described in Sec.~\ref{subsec:The_VC_methos},
allows for obtaining not only the ground-state energy but also correlation
functions of the form $C_{m\mu n\nu}\equiv\langle\hat{L}_{m\mu}^{(k)}\hat{L}_{n\nu}^{(k)}\rangle$.
We now examine the correlation functions obtained from the VC approximation,
focusing on $k=3$, for which $\hat{L}_{n;i\alpha\beta}^{(k=3)}\equiv\hat{\sigma}_{n}^{i}\hat{\sigma}_{n+1}^{\alpha}\hat{\sigma}_{n+2}^{\beta}$.

\begin{figure}
\begin{centering}
\begin{tabular}{lr}
\hskip -2mm
\includegraphics[clip=true,trim=0mm 0mm 0mm 0mm,height=3.4cm]{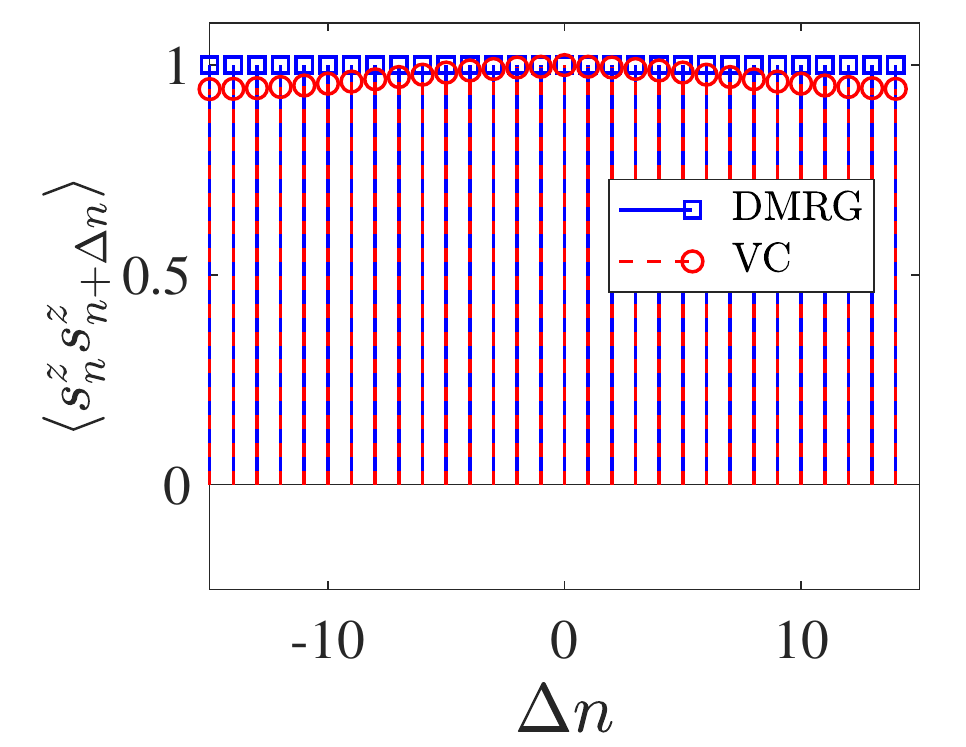}
\llap{\hskip 2mm \parbox[c]{8.5cm}{\vspace{-6.2cm}(a)}}
&
\hskip -2mm
\includegraphics[clip=true,trim=0mm 0mm 0mm 0mm,height=3.4cm]{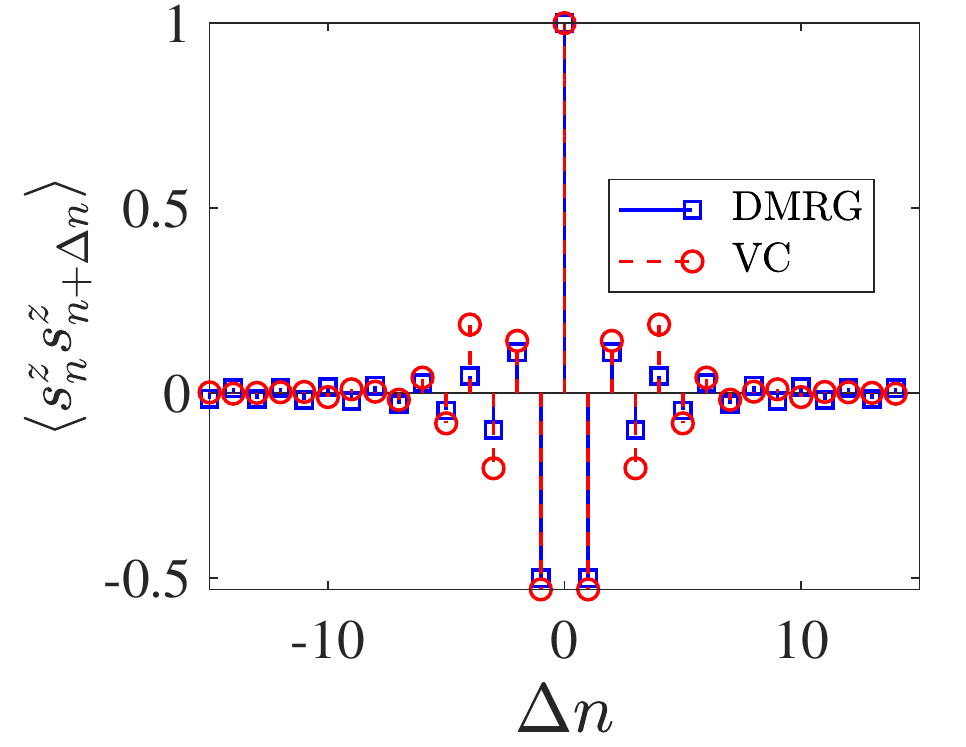}
\llap{\hskip 2mm \parbox[c]{8.5cm}{\vspace{-6.2cm}(b)}}
\\
\includegraphics[clip=true,trim=0mm 0mm 0mm 0mm,height=3.4cm]{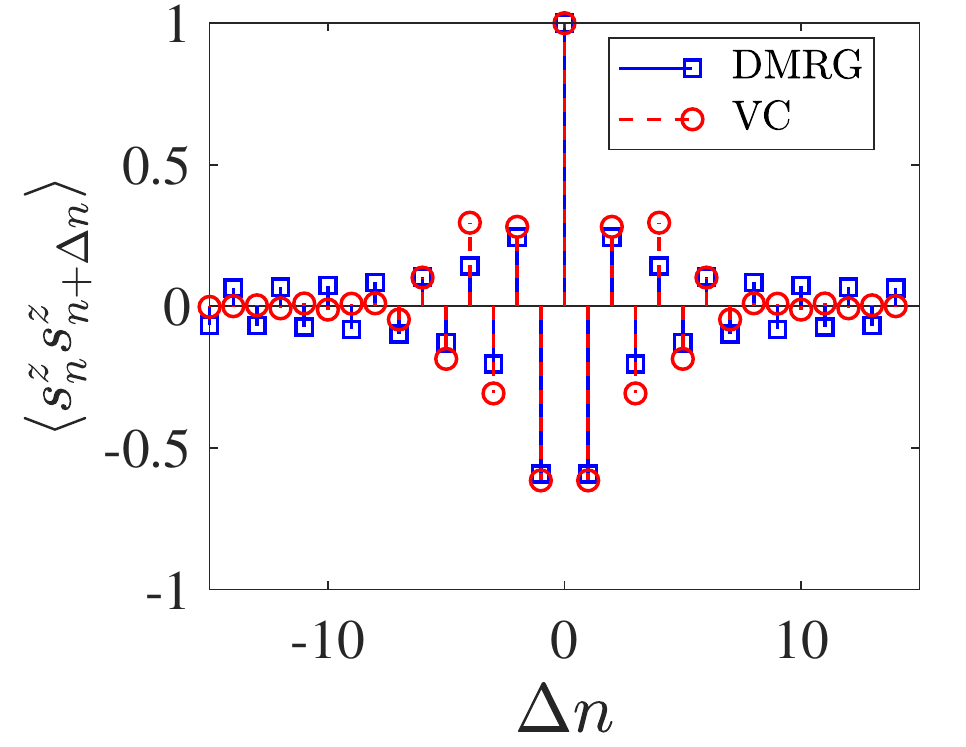}
\llap{\hskip 2mm \parbox[c]{8.5cm}{\vspace{-6.2cm}(c)}}
&
\hskip -2mm
\includegraphics[clip=true,trim=0mm 0mm 0mm 0mm,height=3.4cm]{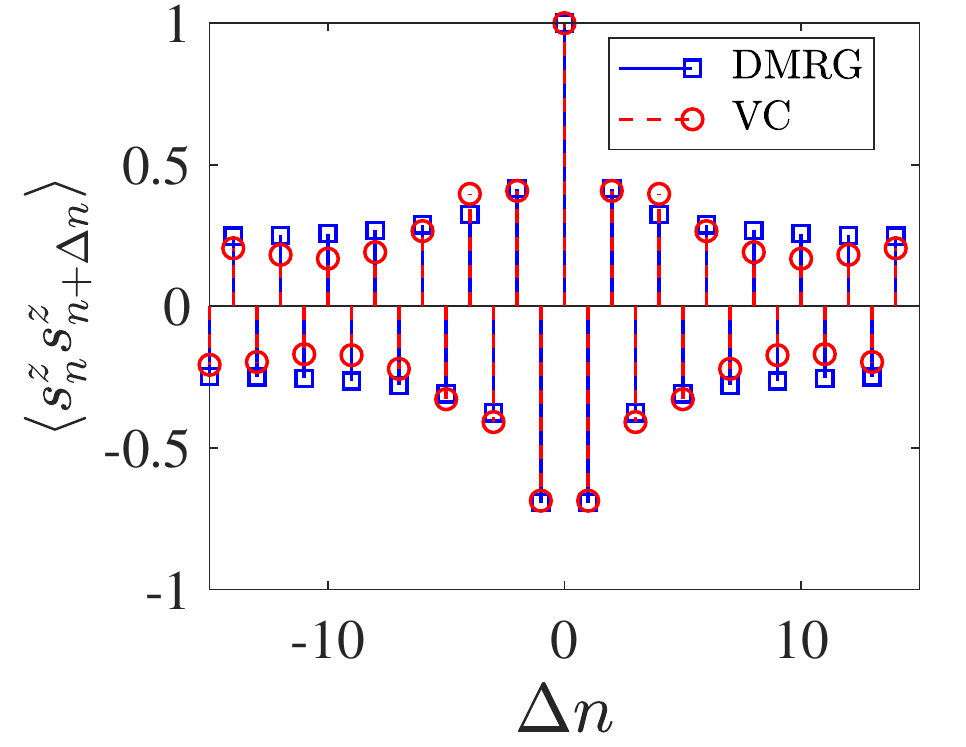}
\llap{\hskip 2mm \parbox[c]{8.5cm}{\vspace{-6.2cm}(d)}}
\end{tabular}
\end{centering}

\caption{Correlation functions for the XXZ model with $N=30$ spins, calculated
to order $k=3$ (red circles) and compared with DMRG (blue squares),
for (a) $J_{z}=-2$, (b) $J_{z}=0.5$, (c) $J_{z}=1,$ and (d) $J_{z}=1.5$.
\label{fig:XXZ_corr}}
\end{figure}

\begin{figure}
\begin{centering}
\begin{tabular}{lr}
\hskip -2mm
\includegraphics[clip=true,trim=0mm 0mm 0mm 0mm,height=3.4cm]{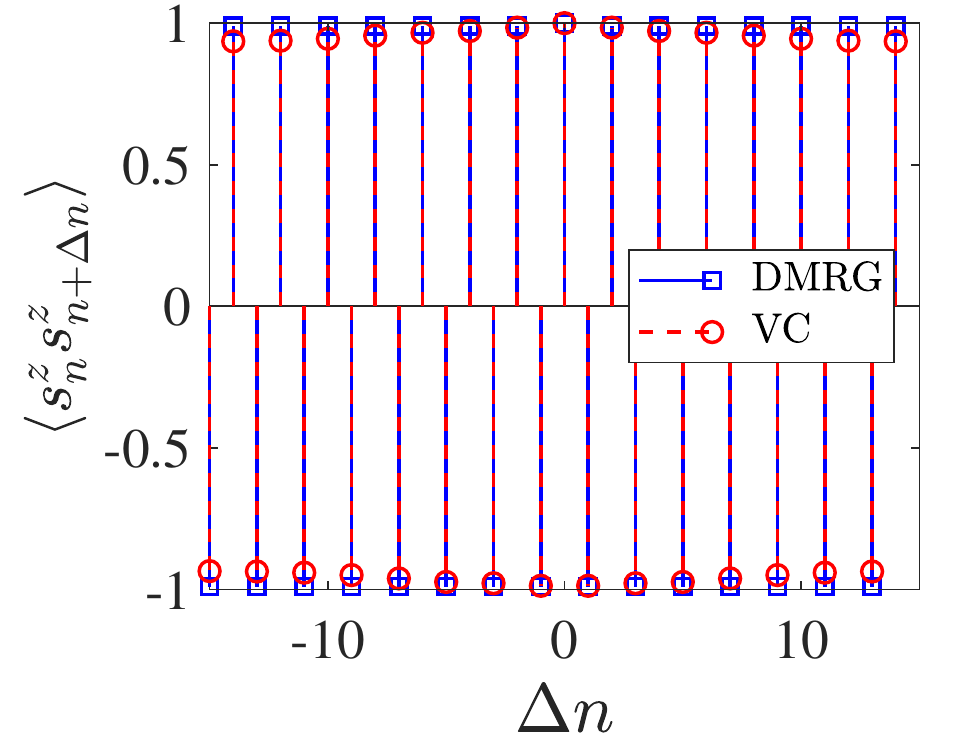}
\llap{\hskip 2mm \parbox[c]{8.5cm}{\vspace{-6.2cm}(a)}}
&
\hskip -2mm
\includegraphics[clip=true,trim=0mm 0mm 0mm 0mm,height=3.4cm]{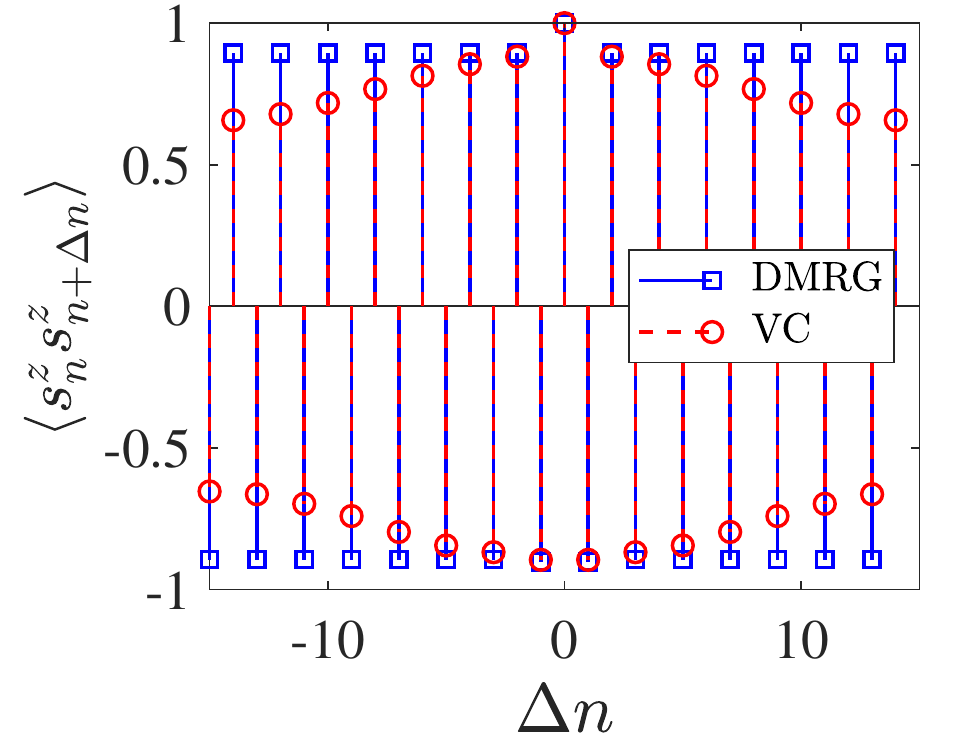}
\llap{\hskip 2mm \parbox[c]{8.5cm}{\vspace{-6.2cm}(b)}}
\\
\hskip -2mm
\includegraphics[clip=true,trim=0mm 0mm 0mm 0mm,height=3.4cm]{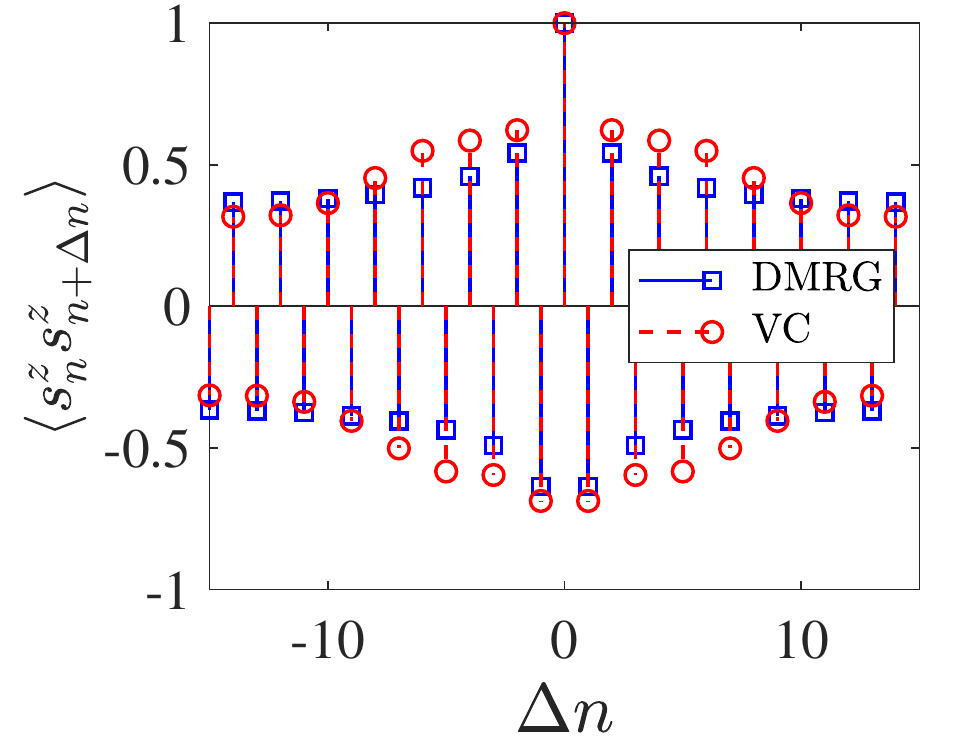}
\llap{\hskip 2mm \parbox[c]{8.5cm}{\vspace{-6.2cm}(c)}}
&
\hskip -2mm
\includegraphics[clip=true,trim=0mm 0mm 0mm 0mm,height=3.4cm]{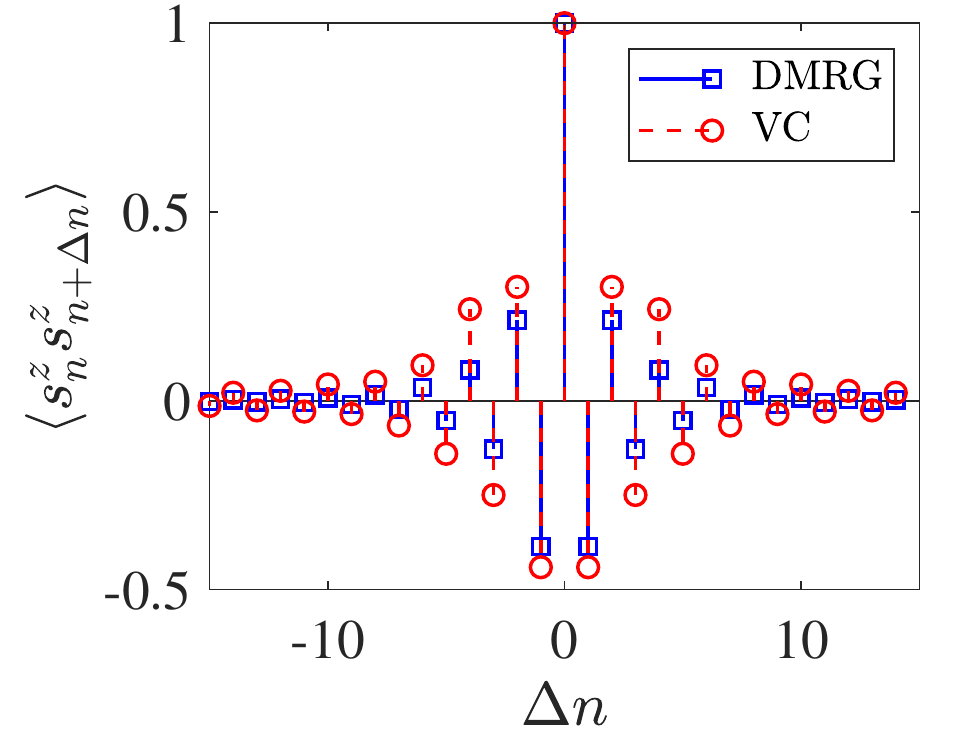}
\llap{\hskip 2mm \parbox[c]{8.5cm}{\vspace{-6.2cm}(d)}}
\end{tabular}
\end{centering}

\caption{Correlation functions for the transverse-field Ising model with $N=30$
spins. Results are shown for the $k=3$ order of the VC approximation
(red circles) and for DMRG (blue squares), for (a) $|\vec{h}|=0.2$,
(b) $|\vec{h}|=0.6$, (c) $|\vec{h}|=1$, and (d) $|\vec{h}|=1.4$.
These correspond to the ordered, critical, and disordered phases of
the Ising model, respectively. \label{fig:Ising_corr}}
\end{figure}

In Fig.~\hyperref[fig:XXZ_corr]{\ref{fig:XXZ_corr}}, we present the
$\langle\hat{\sigma}_{n}^{z}\hat{\sigma}_{n+\Delta n}^{z}\rangle$
correlations for the XXZ model studied in Figs.~\hyperref[fig:E_gs_vs_k]{\ref{fig:E_gs_vs_k}(a,b)},
for several values of $J_{z}$, obtained from the $k=3$ order of
the VC approximation. As before, the results are compared with those
of a DMRG calculation. The VC method captures correctly the qualitative
ferromagnetic {[}Fig.~\hyperref[fig:XXZ_corr]{\ref{fig:XXZ_corr}(a)}{]}
and antiferromagnetic correlations {[}Fig.~\hyperref[fig:XXZ_corr]{\ref{fig:XXZ_corr}(b-d)}{]}.
For small $\Delta n$, good quantitative agreement is observed. However,
that slightly diminishes when increasing $\Delta n$. The same level
of agreement is obtained when examining other kinds of spin-spin correlations
(e.g. $\langle\hat{\sigma}_{m}^{x}\hat{\sigma}_{n}^{x}\rangle$).
Similar conclusions can be drawn from Fig.~\hyperref[fig:Ising_corr]{\ref{fig:Ising_corr}},
which presents the $\langle\sigma_{n}^{z}\sigma_{n+\Delta n}^{z}\rangle$
correlations for the transverse-field Ising model, for several fixed
values of $|\vec{h}|$.

\section{Discussion\label{sec:Discussion}}

We have investigated the variational correlations (VC) approach for
studying the ground state of interacting many-body systems. In this
approach, the elements of a correlation matrix, whose dimension is
linear in system size, serve as the variational parameters, and replace
the density matrix in describing the system. The variational procedure
then relies on using this correlation matrix for obtaining an order-by-order
approximation of the positive semidefiniteness condition of the density
matrix. Since in this variational procedure one relaxes the constraint
on the density matrix rather than over-constraining it, the resulting
energy sets a \emph{lower bound} on the true ground state energy,
similar to the variational 2-RDM method~\citep{Mazziotti2002variational,Zhao2004reduced,Mazziotti2005variational,Mazziotti2006quantum,Mazziotti2007reduced,Barthel2012Solving,Baumgratz2012Lower,Anderson2013second,Verstichel2013Extensive,Mazziotti2016enhanced,Alcoba2018direct,Rubio2019variational}.

The VC approach was tested on several 1d systems of spin 1/2 particles
by comparing its results with those of DMRG, which for 1d systems
is essentially exact. It was demonstrated that the VC approach is
able to produce long-range correlations, as well as to provide a lower
bound on the ground-state energy that converges to the exact result
as the order of approximation is increased. Interestingly, the VC
approximation becomes better in the presence of disorder. The fact
that the VC approach provides a lower bound could be used together
with a conventional variational ansatz (e.g. mean field) to bound
the ground state energy from both above and below.

In 1d, the VC method offers no advantage over DMRG in terms of computational
complexity, as both of them scale polynomially~\footnote{At each step of the minimization procedure, Eq.~(\ref{eq:VC_opt_prob}),
one has to diagonalize the correlation matrix, $\mathcal{M}^{(k)}$,
whose dimension scales linearly with the system size, $N$. The overall
number of variational parameters, $x$, scales either as $N^{2}$
or as $N$, the latter case corresponding to translationally-invariant
systems.}, and for a gapped system DMRG scales linearly. In 2d, however, DMRG
scales exponentially with the width of the system~\citep{Stoudenmire2012Studying},
while the VC method remains polynomial. It will, therefore, be interesting
to examine the VC approach when applied to models in 2d, where it
has the potential to address some outstanding challenges, both in
condensed matter and in cold atoms systems. Since the VC approximation
can be formulated as an Semidefinite Programming~\citep{Vandenberghe1996semidefinite}
(SDP) problem, recent advances~\citep{Yurtsever2017sketchy} in large-scale
SDP algorithms could help to achieve this goal, as well as to attend
higher-approximation orders. Interestingly, it has been suggested
that SDP is one of a few problems that could acquire a speedup from
the introduction of \emph{Noisy Intermediate-Scale Quantum} (NISQ)
technology~\citep{Preskill_2018}, through a recently-introduced
algorithm dubbed Quantum Semidefinite Programming~\citep{Brandao2017quantum,VanApeldoorn2017quantum,Brandao2019faster}.

As explained in Sec.~\ref{subsec:approx_PSD}, the VC approach is
based on relaxing the condition of Eq.~(\ref{eq:rho_psd_equiv_Var_o}),
by requiring it is obeyed for only a subset of operators, denoted
$\hat{O}^{(k)}$, of range $k$-local operators. Clearly, one can
choose a different subset of operators, amounting to a different way
of approximating the ground state. Two such examples are the approximations
employed in Refs.~\citep{Mazziotti2002variational,Zhao2004reduced,Mazziotti2005variational,Mazziotti2006quantum,Mazziotti2007reduced,Baumgratz2012Lower,Anderson2013second,Verstichel2013Extensive,Mazziotti2016enhanced,Alcoba2018direct,Rubio2019variational}
and Ref.~\citep{Barthel2012Solving}. It will therefore be interesting to study the effect of choosing
different subsets of operators in Eq.~(\ref{eq:O_k}), and the physical
meaning of their resulting approximations. In particular, it is reasonable
to assume that the optimal choice could depend on the properties of
the ground-state being targeted.

Finally, while the VC method targets the ground-state, a possible
extension might be to study excited states. This could possibly be
achieved using the fact that the variance of the energy, $\langle H^{2}\rangle-\langle H\rangle^{2}$,
which is a linear function of the correlation matrix, is a non-negative
quantity that vanishes only for eigenstates. Minimizing it, while
constraining $\langle H\rangle$ to lie withing a narrow window, could
potentially yield an approximation for the correlation functions at
finite energy density.

\section*{Acknowledgments}

We have benefited from discussions with Y. Baum, O. Motrunich, E.
P. L. van Nieuwenburg, K. Slagel, and C. D. White. This research was
supported by the Institute of Quantum Information and Matter, an NSF
Frontier center funded by the Gordon and Betty Moore Foundation, the
Packard Foundation, and the Simons foundation. AH acknowledges support
from the Walter Burke Institute for Theoretical Physics at Caltech.
RK acknowledges funding provided by the Office of Naval Research (Award
N00014-17-1-2146) and the Army Research Office (Award W911NF121054).

\appendix

\section{Implementation\label{sec:Implementation}}

To numerically solve the minimization problem of Eq.~(\ref{eq:VC_opt_prob}),
we first formulate it as a semidefinite programming (SDP) problem~\citep{Vandenberghe1996semidefinite}.
To this end we write the correlation matrix, defined in Eqs.~(\ref{eq:M_k})
and~(\ref{eq:B_C_def}), as
\begin{equation}
\mathcal{M}_{m\mu;n\nu}^{(k)}=C_{m\mu;n\nu}-b_{m\mu}b_{n\nu,}
\end{equation}
and use the fact that the condition $\mathcal{M}^{(k)}\succeq0$ is
equivalent to
\begin{equation}
X=\begin{pmatrix}C & b\\
b^{{\rm T}} & 1
\end{pmatrix}\succeq0.
\end{equation}

The objective function, $E=\langle H\rangle$ , can now be written
as a linear function of the positive-semidefinite matrix, $X$. Notice
also that the elements of $C_{m\mu;n\nu}=\langle\hat{L}_{m\mu}^{(k)}\hat{L}_{n\nu}^{(k)}\rangle$
for $|n-m|<k$ are not independent, and can be expressed as linear
functions of the other matrix elements. This therefore constitutes
a SDP problem which we then solve using the CVX package for specifying
and solving convex programs~\citep{CVX}, with the MOSEK interior-point
solver~\citep{mosek}.

For example, for $k=2,$ one has 
\begin{equation}
b_{i\alpha}^{n}=\langle\sigma_{n}^{i}\sigma_{n+1}^{\alpha}\rangle\hspace{1em};\hspace{1em}C_{i\alpha j\beta}^{mn}=\langle\sigma_{m}^{i}\sigma_{m+1}^{\alpha}\sigma_{n}^{j}\sigma_{n+1}^{\beta}\rangle.
\end{equation}
The energy can be written as 

\begin{equation}
E=\langle\hat{H}\rangle=\sum_{mi\alpha}J_{mi\alpha}b_{mi\alpha}=\sum_{mi\alpha}J_{mi\alpha}X_{12N+1;mi\alpha},
\end{equation}
and the matrix $X$ obeys the following linear constraints\begin{widetext}
\begin{equation}
\begin{aligned} & X_{mi\alpha;m+1,j\beta}={\rm Re}(\kappa_{\alpha j\gamma})\kappa_{\beta l0}\kappa_{mi\gamma;m+2,l0}+{\rm Re(}\kappa_{\alpha j\gamma})\delta_{\beta0}X_{12N+1;mi\gamma,}\\
 & X_{mi\alpha;mj\beta}={\rm Re}(\kappa_{ijl}\kappa_{\alpha\beta\delta})X_{12N+1;ml\delta}+{\rm Re}(\kappa_{ij0}\kappa_{\alpha\beta k})X_{12N+1;m+1,k0}+\delta_{ij}\delta_{\alpha\beta},
\end{aligned}
\end{equation}
\end{widetext}where the tensor $\kappa$ is defined by $\sigma^{\alpha}\sigma^{\beta}=\kappa_{\alpha\beta\gamma}\sigma^{\gamma}$.
Finally, when solving for the ground-state energy and correlation
functions of Figs.~\ref{fig:E_gs_vs_k}-\ref{fig:Ising_corr} we
first use the translational invariance of the model in order to reduce
the number of parameters as explained in Sec.~\ref{subsec:trans_inv}.
The minimization problem is then solved using an interior-point algorithm
for nonlinear optimization problems implemented by MATLAB.

\bibliographystyle{apsrev4-1}
\bibliography{References}

\begin{thebibliography}{34}%
\makeatletter
\providecommand \@ifxundefined [1]{%
 \@ifx{#1\undefined}
}%
\providecommand \@ifnum [1]{%
 \ifnum #1\expandafter \@firstoftwo
 \else \expandafter \@secondoftwo
 \fi
}%
\providecommand \@ifx [1]{%
 \ifx #1\expandafter \@firstoftwo
 \else \expandafter \@secondoftwo
 \fi
}%
\providecommand \natexlab [1]{#1}%
\providecommand \enquote  [1]{``#1''}%
\providecommand \bibnamefont  [1]{#1}%
\providecommand \bibfnamefont [1]{#1}%
\providecommand \citenamefont [1]{#1}%
\providecommand \href@noop [0]{\@secondoftwo}%
\providecommand \href [0]{\begingroup \@sanitize@url \@href}%
\providecommand \@href[1]{\@@startlink{#1}\@@href}%
\providecommand \@@href[1]{\endgroup#1\@@endlink}%
\providecommand \@sanitize@url [0]{\catcode `\\12\catcode `\$12\catcode
  `\&12\catcode `\#12\catcode `\^12\catcode `\_12\catcode `\%12\relax}%
\providecommand \@@startlink[1]{}%
\providecommand \@@endlink[0]{}%
\providecommand \url  [0]{\begingroup\@sanitize@url \@url }%
\providecommand \@url [1]{\endgroup\@href {#1}{\urlprefix }}%
\providecommand \urlprefix  [0]{URL }%
\providecommand \Eprint [0]{\href }%
\providecommand \doibase [0]{http://dx.doi.org/}%
\providecommand \selectlanguage [0]{\@gobble}%
\providecommand \bibinfo  [0]{\@secondoftwo}%
\providecommand \bibfield  [0]{\@secondoftwo}%
\providecommand \translation [1]{[#1]}%
\providecommand \BibitemOpen [0]{}%
\providecommand \bibitemStop [0]{}%
\providecommand \bibitemNoStop [0]{.\EOS\space}%
\providecommand \EOS [0]{\spacefactor3000\relax}%
\providecommand \BibitemShut  [1]{\csname bibitem#1\endcsname}%
\let\auto@bib@innerbib\@empty
\bibitem [{\citenamefont {White}(1992)}]{White1992PRL}%
  \BibitemOpen
  \bibfield  {author} {\bibinfo {author} {\bibfnamefont {S.~R.}\ \bibnamefont
  {White}},\ }\href {\doibase 10.1103/PhysRevLett.69.2863} {\bibfield
  {journal} {\bibinfo  {journal} {Phys. Rev. Lett.}\ }\textbf {\bibinfo
  {volume} {69}},\ \bibinfo {pages} {2863} (\bibinfo {year}
  {1992})}\BibitemShut {NoStop}%
\bibitem [{\citenamefont {White}(1993)}]{White1993PRB}%
  \BibitemOpen
  \bibfield  {author} {\bibinfo {author} {\bibfnamefont {S.~R.}\ \bibnamefont
  {White}},\ }\href {\doibase 10.1103/PhysRevB.48.10345} {\bibfield  {journal}
  {\bibinfo  {journal} {Phys. Rev. B}\ }\textbf {\bibinfo {volume} {48}},\
  \bibinfo {pages} {10345} (\bibinfo {year} {1993})}\BibitemShut {NoStop}%
\bibitem [{\citenamefont {Schollw\"ock}(2005)}]{Schollwock2005TheDensity}%
  \BibitemOpen
  \bibfield  {author} {\bibinfo {author} {\bibfnamefont {U.}~\bibnamefont
  {Schollw\"ock}},\ }\href {\doibase 10.1103/RevModPhys.77.259} {\bibfield
  {journal} {\bibinfo  {journal} {Rev. Mod. Phys.}\ }\textbf {\bibinfo {volume}
  {77}},\ \bibinfo {pages} {259} (\bibinfo {year} {2005})}\BibitemShut
  {NoStop}%
\bibitem [{\citenamefont {Gubernatis}\ \emph {et~al.}(2016)\citenamefont
  {Gubernatis}, \citenamefont {Kawashima},\ and\ \citenamefont
  {Werner}}]{Gubernatis2016quantum}%
  \BibitemOpen
  \bibfield  {author} {\bibinfo {author} {\bibfnamefont {J.}~\bibnamefont
  {Gubernatis}}, \bibinfo {author} {\bibfnamefont {N.}~\bibnamefont
  {Kawashima}}, \ and\ \bibinfo {author} {\bibfnamefont {P.}~\bibnamefont
  {Werner}},\ }\href@noop {} {\emph {\bibinfo {title} {Quantum Monte Carlo
  Methods}}}\ (\bibinfo  {publisher} {Cambridge University Press},\ \bibinfo
  {year} {2016})\BibitemShut {NoStop}%
\bibitem [{\citenamefont {Loh}\ \emph {et~al.}(1990)\citenamefont {Loh},
  \citenamefont {Gubernatis}, \citenamefont {Scalettar}, \citenamefont {White},
  \citenamefont {Scalapino},\ and\ \citenamefont {Sugar}}]{Loh1990sign}%
  \BibitemOpen
  \bibfield  {author} {\bibinfo {author} {\bibfnamefont {E.~Y.}\ \bibnamefont
  {Loh}}, \bibinfo {author} {\bibfnamefont {J.~E.}\ \bibnamefont {Gubernatis}},
  \bibinfo {author} {\bibfnamefont {R.~T.}\ \bibnamefont {Scalettar}}, \bibinfo
  {author} {\bibfnamefont {S.~R.}\ \bibnamefont {White}}, \bibinfo {author}
  {\bibfnamefont {D.~J.}\ \bibnamefont {Scalapino}}, \ and\ \bibinfo {author}
  {\bibfnamefont {R.~L.}\ \bibnamefont {Sugar}},\ }\href {\doibase
  10.1103/PhysRevB.41.9301} {\bibfield  {journal} {\bibinfo  {journal} {Phys.
  Rev. B}\ }\textbf {\bibinfo {volume} {41}},\ \bibinfo {pages} {9301}
  (\bibinfo {year} {1990})}\BibitemShut {NoStop}%
\bibitem [{\citenamefont {Carleo}\ and\ \citenamefont
  {Troyer}(2017)}]{Carleo2017solving}%
  \BibitemOpen
  \bibfield  {author} {\bibinfo {author} {\bibfnamefont {G.}~\bibnamefont
  {Carleo}}\ and\ \bibinfo {author} {\bibfnamefont {M.}~\bibnamefont
  {Troyer}},\ }\href {\doibase 10.1126/science.aag2302} {\bibfield  {journal}
  {\bibinfo  {journal} {Science}\ }\textbf {\bibinfo {volume} {355}},\ \bibinfo
  {pages} {602–606} (\bibinfo {year} {2017})}\BibitemShut {NoStop}%
\bibitem [{\citenamefont {Melko}\ \emph {et~al.}(2019)\citenamefont {Melko},
  \citenamefont {Carleo}, \citenamefont {Carrasquilla},\ and\ \citenamefont
  {Cirac}}]{Melko2019restricted}%
  \BibitemOpen
  \bibfield  {author} {\bibinfo {author} {\bibfnamefont {R.~G.}\ \bibnamefont
  {Melko}}, \bibinfo {author} {\bibfnamefont {G.}~\bibnamefont {Carleo}},
  \bibinfo {author} {\bibfnamefont {J.}~\bibnamefont {Carrasquilla}}, \ and\
  \bibinfo {author} {\bibfnamefont {J.~I.}\ \bibnamefont {Cirac}},\ }\href
  {https://www.nature.com/articles/s41567-019-0545-1} {\bibfield  {journal}
  {\bibinfo  {journal} {Nature Physics}\ ,\ \bibinfo {pages} {1}} (\bibinfo
  {year} {2019})}\BibitemShut {NoStop}%
\bibitem [{\citenamefont {Sharir}\ \emph {et~al.}(2019)\citenamefont {Sharir},
  \citenamefont {Levine}, \citenamefont {Wies}, \citenamefont {Carleo},\ and\
  \citenamefont {Shashua}}]{Sharir2019deep}%
  \BibitemOpen
  \bibfield  {author} {\bibinfo {author} {\bibfnamefont {O.}~\bibnamefont
  {Sharir}}, \bibinfo {author} {\bibfnamefont {Y.}~\bibnamefont {Levine}},
  \bibinfo {author} {\bibfnamefont {N.}~\bibnamefont {Wies}}, \bibinfo {author}
  {\bibfnamefont {G.}~\bibnamefont {Carleo}}, \ and\ \bibinfo {author}
  {\bibfnamefont {A.}~\bibnamefont {Shashua}},\ }\href@noop {} {\enquote
  {\bibinfo {title} {Deep autoregressive models for the efficient variational
  simulation of many-body quantum systems},}\ } (\bibinfo {year} {2019}),\
  \Eprint {http://arxiv.org/abs/1902.04057} {arXiv:1902.04057
  [cond-mat.dis-nn]} \BibitemShut {NoStop}%
\bibitem [{\citenamefont {Mazziotti}(2002)}]{Mazziotti2002variational}%
  \BibitemOpen
  \bibfield  {author} {\bibinfo {author} {\bibfnamefont {D.~A.}\ \bibnamefont
  {Mazziotti}},\ }\href {\doibase 10.1103/PhysRevA.65.062511} {\bibfield
  {journal} {\bibinfo  {journal} {Phys. Rev. A}\ }\textbf {\bibinfo {volume}
  {65}},\ \bibinfo {pages} {062511} (\bibinfo {year} {2002})}\BibitemShut
  {NoStop}%
\bibitem [{\citenamefont {Zhao}\ \emph {et~al.}(2004)\citenamefont {Zhao},
  \citenamefont {Braams}, \citenamefont {Fukuda}, \citenamefont {Overton},\
  and\ \citenamefont {Percus}}]{Zhao2004reduced}%
  \BibitemOpen
  \bibfield  {author} {\bibinfo {author} {\bibfnamefont {Z.}~\bibnamefont
  {Zhao}}, \bibinfo {author} {\bibfnamefont {B.~J.}\ \bibnamefont {Braams}},
  \bibinfo {author} {\bibfnamefont {M.}~\bibnamefont {Fukuda}}, \bibinfo
  {author} {\bibfnamefont {M.~L.}\ \bibnamefont {Overton}}, \ and\ \bibinfo
  {author} {\bibfnamefont {J.~K.}\ \bibnamefont {Percus}},\ }\href@noop {}
  {\bibfield  {journal} {\bibinfo  {journal} {The Journal of chemical physics}\
  }\textbf {\bibinfo {volume} {120}},\ \bibinfo {pages} {2095} (\bibinfo {year}
  {2004})}\BibitemShut {NoStop}%
\bibitem [{\citenamefont {Mazziotti}(2005)}]{Mazziotti2005variational}%
  \BibitemOpen
  \bibfield  {author} {\bibinfo {author} {\bibfnamefont {D.~A.}\ \bibnamefont
  {Mazziotti}},\ }\href {\doibase 10.1103/PhysRevA.72.032510} {\bibfield
  {journal} {\bibinfo  {journal} {Phys. Rev. A}\ }\textbf {\bibinfo {volume}
  {72}},\ \bibinfo {pages} {032510} (\bibinfo {year} {2005})}\BibitemShut
  {NoStop}%
\bibitem [{\citenamefont {Mazziotti}(2006)}]{Mazziotti2006quantum}%
  \BibitemOpen
  \bibfield  {author} {\bibinfo {author} {\bibfnamefont {D.~A.}\ \bibnamefont
  {Mazziotti}},\ }\href@noop {} {\bibfield  {journal} {\bibinfo  {journal}
  {Accounts of chemical research}\ }\textbf {\bibinfo {volume} {39}},\ \bibinfo
  {pages} {207} (\bibinfo {year} {2006})}\BibitemShut {NoStop}%
\bibitem [{\citenamefont {Mazziotti}(2007)}]{Mazziotti2007reduced}%
  \BibitemOpen
  \bibfield  {author} {\bibinfo {author} {\bibfnamefont {D.~A.}\ \bibnamefont
  {Mazziotti}},\ }\href@noop {} {\emph {\bibinfo {title}
  {Reduced-density-matrix mechanics: with applications to many-electron atoms
  and molecules}}},\ Vol.\ \bibinfo {volume} {134}\ (\bibinfo  {publisher}
  {Wiley Online Library},\ \bibinfo {year} {2007})\BibitemShut {NoStop}%
\bibitem [{\citenamefont {Barthel}\ and\ \citenamefont
  {H\"ubener}(2012)}]{Barthel2012Solving}%
  \BibitemOpen
  \bibfield  {author} {\bibinfo {author} {\bibfnamefont {T.}~\bibnamefont
  {Barthel}}\ and\ \bibinfo {author} {\bibfnamefont {R.}~\bibnamefont
  {H\"ubener}},\ }\href {\doibase 10.1103/PhysRevLett.108.200404} {\bibfield
  {journal} {\bibinfo  {journal} {Phys. Rev. Lett.}\ }\textbf {\bibinfo
  {volume} {108}},\ \bibinfo {pages} {200404} (\bibinfo {year}
  {2012})}\BibitemShut {NoStop}%
\bibitem [{\citenamefont {Baumgratz}\ and\ \citenamefont
  {Plenio}(2012)}]{Baumgratz2012Lower}%
  \BibitemOpen
  \bibfield  {author} {\bibinfo {author} {\bibfnamefont {T.}~\bibnamefont
  {Baumgratz}}\ and\ \bibinfo {author} {\bibfnamefont {M.~B.}\ \bibnamefont
  {Plenio}},\ }\href {\doibase 10.1088/1367-2630/14/2/023027} {\bibfield
  {journal} {\bibinfo  {journal} {New Journal of Physics}\ }\textbf {\bibinfo
  {volume} {14}},\ \bibinfo {pages} {023027} (\bibinfo {year}
  {2012})}\BibitemShut {NoStop}%
\bibitem [{\citenamefont {Anderson}\ \emph {et~al.}(2013)\citenamefont
  {Anderson}, \citenamefont {Nakata}, \citenamefont {Igarashi}, \citenamefont
  {Fujisawa},\ and\ \citenamefont {Yamashita}}]{Anderson2013second}%
  \BibitemOpen
  \bibfield  {author} {\bibinfo {author} {\bibfnamefont {J.~S.}\ \bibnamefont
  {Anderson}}, \bibinfo {author} {\bibfnamefont {M.}~\bibnamefont {Nakata}},
  \bibinfo {author} {\bibfnamefont {R.}~\bibnamefont {Igarashi}}, \bibinfo
  {author} {\bibfnamefont {K.}~\bibnamefont {Fujisawa}}, \ and\ \bibinfo
  {author} {\bibfnamefont {M.}~\bibnamefont {Yamashita}},\ }\href@noop {}
  {\bibfield  {journal} {\bibinfo  {journal} {Computational and Theoretical
  Chemistry}\ }\textbf {\bibinfo {volume} {1003}},\ \bibinfo {pages} {22}
  (\bibinfo {year} {2013})}\BibitemShut {NoStop}%
\bibitem [{\citenamefont {Verstichel}\ \emph {et~al.}(2013)\citenamefont
  {Verstichel}, \citenamefont {van Aggelen}, \citenamefont {Poelmans},
  \citenamefont {Wouters},\ and\ \citenamefont
  {Van~Neck}}]{Verstichel2013Extensive}%
  \BibitemOpen
  \bibfield  {author} {\bibinfo {author} {\bibfnamefont {B.}~\bibnamefont
  {Verstichel}}, \bibinfo {author} {\bibfnamefont {H.}~\bibnamefont {van
  Aggelen}}, \bibinfo {author} {\bibfnamefont {W.}~\bibnamefont {Poelmans}},
  \bibinfo {author} {\bibfnamefont {S.}~\bibnamefont {Wouters}}, \ and\
  \bibinfo {author} {\bibfnamefont {D.}~\bibnamefont {Van~Neck}},\ }\href
  {\doibase 10.1016/j.comptc.2012.09.014} {\bibfield  {journal} {\bibinfo
  {journal} {Computational and Theoretical Chemistry}\ }\textbf {\bibinfo
  {volume} {1003}},\ \bibinfo {pages} {12–21} (\bibinfo {year}
  {2013})}\BibitemShut {NoStop}%
\bibitem [{\citenamefont {Mazziotti}(2016)}]{Mazziotti2016enhanced}%
  \BibitemOpen
  \bibfield  {author} {\bibinfo {author} {\bibfnamefont {D.~A.}\ \bibnamefont
  {Mazziotti}},\ }\href {\doibase 10.1103/PhysRevLett.117.153001} {\bibfield
  {journal} {\bibinfo  {journal} {Phys. Rev. Lett.}\ }\textbf {\bibinfo
  {volume} {117}},\ \bibinfo {pages} {153001} (\bibinfo {year}
  {2016})}\BibitemShut {NoStop}%
\bibitem [{\citenamefont {Alcoba}\ \emph {et~al.}(2018)\citenamefont {Alcoba},
  \citenamefont {Torre}, \citenamefont {Lain}, \citenamefont {Massaccesi},
  \citenamefont {Ona}, \citenamefont {Honor{\'e}}, \citenamefont {Poelmans},
  \citenamefont {Van~Neck}, \citenamefont {Bultinck},\ and\ \citenamefont
  {De~Baerdemacker}}]{Alcoba2018direct}%
  \BibitemOpen
  \bibfield  {author} {\bibinfo {author} {\bibfnamefont {D.~R.}\ \bibnamefont
  {Alcoba}}, \bibinfo {author} {\bibfnamefont {A.}~\bibnamefont {Torre}},
  \bibinfo {author} {\bibfnamefont {L.}~\bibnamefont {Lain}}, \bibinfo {author}
  {\bibfnamefont {G.~E.}\ \bibnamefont {Massaccesi}}, \bibinfo {author}
  {\bibfnamefont {O.~B.}\ \bibnamefont {Ona}}, \bibinfo {author} {\bibfnamefont
  {E.~M.}\ \bibnamefont {Honor{\'e}}}, \bibinfo {author} {\bibfnamefont
  {W.}~\bibnamefont {Poelmans}}, \bibinfo {author} {\bibfnamefont
  {D.}~\bibnamefont {Van~Neck}}, \bibinfo {author} {\bibfnamefont
  {P.}~\bibnamefont {Bultinck}}, \ and\ \bibinfo {author} {\bibfnamefont
  {S.}~\bibnamefont {De~Baerdemacker}},\ }\href@noop {} {\bibfield  {journal}
  {\bibinfo  {journal} {The Journal of chemical physics}\ }\textbf {\bibinfo
  {volume} {148}},\ \bibinfo {pages} {024105} (\bibinfo {year}
  {2018})}\BibitemShut {NoStop}%
\bibitem [{\citenamefont {Rubio-Garc{\'\i}a}\ \emph {et~al.}(2019)\citenamefont
  {Rubio-Garc{\'\i}a}, \citenamefont {Dukelsky}, \citenamefont {Alcoba},
  \citenamefont {Capuzzi}, \citenamefont {O{\~n}a}, \citenamefont {R{\'\i}os},
  \citenamefont {Torre},\ and\ \citenamefont {Lain}}]{Rubio2019variational}%
  \BibitemOpen
  \bibfield  {author} {\bibinfo {author} {\bibfnamefont {A.}~\bibnamefont
  {Rubio-Garc{\'\i}a}}, \bibinfo {author} {\bibfnamefont {J.}~\bibnamefont
  {Dukelsky}}, \bibinfo {author} {\bibfnamefont {D.}~\bibnamefont {Alcoba}},
  \bibinfo {author} {\bibfnamefont {P.}~\bibnamefont {Capuzzi}}, \bibinfo
  {author} {\bibfnamefont {O.}~\bibnamefont {O{\~n}a}}, \bibinfo {author}
  {\bibfnamefont {E.}~\bibnamefont {R{\'\i}os}}, \bibinfo {author}
  {\bibfnamefont {A.}~\bibnamefont {Torre}}, \ and\ \bibinfo {author}
  {\bibfnamefont {L.}~\bibnamefont {Lain}},\ }\href@noop {} {\bibfield
  {journal} {\bibinfo  {journal} {The Journal of chemical physics}\ }\textbf
  {\bibinfo {volume} {151}},\ \bibinfo {pages} {154104} (\bibinfo {year}
  {2019})}\BibitemShut {NoStop}%
\bibitem [{\citenamefont {Qi}\ and\ \citenamefont
  {Ranard}(2019)}]{Qi2019Determining}%
  \BibitemOpen
  \bibfield  {author} {\bibinfo {author} {\bibfnamefont {X.-L.}\ \bibnamefont
  {Qi}}\ and\ \bibinfo {author} {\bibfnamefont {D.}~\bibnamefont {Ranard}},\
  }\href {\doibase 10.22331/q-2019-07-08-159} {\bibfield  {journal} {\bibinfo
  {journal} {Quantum}\ }\textbf {\bibinfo {volume} {3}},\ \bibinfo {pages}
  {159} (\bibinfo {year} {2019})}\BibitemShut {NoStop}%
\bibitem [{\citenamefont {Mazziotti}\ and\ \citenamefont
  {Erdahl}(2001)}]{Mazziotti2001Uncertainty}%
  \BibitemOpen
  \bibfield  {author} {\bibinfo {author} {\bibfnamefont {D.~A.}\ \bibnamefont
  {Mazziotti}}\ and\ \bibinfo {author} {\bibfnamefont {R.~M.}\ \bibnamefont
  {Erdahl}},\ }\href {\doibase 10.1103/PhysRevA.63.042113} {\bibfield
  {journal} {\bibinfo  {journal} {Phys. Rev. A}\ }\textbf {\bibinfo {volume}
  {63}},\ \bibinfo {pages} {042113} (\bibinfo {year} {2001})}\BibitemShut
  {NoStop}%
\bibitem [{\citenamefont {Suzuki}\ \emph {et~al.}(2012)\citenamefont {Suzuki},
  \citenamefont {Inoue},\ and\ \citenamefont
  {Chakrabarti}}]{Suzuki2012quantum}%
  \BibitemOpen
  \bibfield  {author} {\bibinfo {author} {\bibfnamefont {S.}~\bibnamefont
  {Suzuki}}, \bibinfo {author} {\bibfnamefont {J.-i.}\ \bibnamefont {Inoue}}, \
  and\ \bibinfo {author} {\bibfnamefont {B.~K.}\ \bibnamefont {Chakrabarti}},\
  }\href@noop {} {\emph {\bibinfo {title} {Quantum Ising phases and transitions
  in transverse Ising models}}},\ Vol.\ \bibinfo {volume} {862}\ (\bibinfo
  {publisher} {Springer},\ \bibinfo {year} {2012})\BibitemShut {NoStop}%
\bibitem [{ITe()}]{ITensor}%
  \BibitemOpen
  \href {http://itensor.org} {\bibinfo  {journal} {\mbox{ITensor Library}
  (version 3.0.0) http://itensor.org}\ }\BibitemShut {NoStop}%
\bibitem [{Note1()}]{Note1}%
  \BibitemOpen
\bibfield  {journal} {  }\bibinfo {note} {At each step of the minimization
  procedure, Eq.~(\ref {eq:VC_opt_prob}), one has to diagonalize the
  correlation matrix, $\protect \mathcal {M}^{(k)}$, whose dimension scales
  linearly with the system size, $N$. The overall number of variational
  parameters, $x$, scales either as $N^{2}$ or as $N$, the latter case
  corresponding to translationally-invariant systems.}\BibitemShut {Stop}%
\bibitem [{\citenamefont {Stoudenmire}\ and\ \citenamefont
  {White}(2012)}]{Stoudenmire2012Studying}%
  \BibitemOpen
  \bibfield  {author} {\bibinfo {author} {\bibfnamefont {E.}~\bibnamefont
  {Stoudenmire}}\ and\ \bibinfo {author} {\bibfnamefont {S.~R.}\ \bibnamefont
  {White}},\ }\href {\doibase 10.1146/annurev-conmatphys-020911-125018}
  {\bibfield  {journal} {\bibinfo  {journal} {Annu. Rev. Condens. Matter
  Phys.}\ }\textbf {\bibinfo {volume} {3}},\ \bibinfo {pages} {111–128}
  (\bibinfo {year} {2012})}\BibitemShut {NoStop}%
\bibitem [{\citenamefont {Vandenberghe}\ and\ \citenamefont
  {Boyd}(1996)}]{Vandenberghe1996semidefinite}%
  \BibitemOpen
  \bibfield  {author} {\bibinfo {author} {\bibfnamefont {L.}~\bibnamefont
  {Vandenberghe}}\ and\ \bibinfo {author} {\bibfnamefont {S.}~\bibnamefont
  {Boyd}},\ }\href {https://epubs.siam.org/doi/abs/10.1137/1038003} {\bibfield
  {journal} {\bibinfo  {journal} {SIAM review}\ }\textbf {\bibinfo {volume}
  {38}},\ \bibinfo {pages} {49} (\bibinfo {year} {1996})}\BibitemShut {NoStop}%
\bibitem [{\citenamefont {Yurtsever}\ \emph {et~al.}(2017)\citenamefont
  {Yurtsever}, \citenamefont {Udell}, \citenamefont {Tropp},\ and\
  \citenamefont {Cevher}}]{Yurtsever2017sketchy}%
  \BibitemOpen
  \bibfield  {author} {\bibinfo {author} {\bibfnamefont {A.}~\bibnamefont
  {Yurtsever}}, \bibinfo {author} {\bibfnamefont {M.}~\bibnamefont {Udell}},
  \bibinfo {author} {\bibfnamefont {J.~A.}\ \bibnamefont {Tropp}}, \ and\
  \bibinfo {author} {\bibfnamefont {V.}~\bibnamefont {Cevher}},\ }\href@noop {}
  {\enquote {\bibinfo {title} {Sketchy decisions: Convex low-rank matrix
  optimization with optimal storage},}\ } (\bibinfo {year} {2017}),\ \Eprint
  {http://arxiv.org/abs/1702.06838} {arXiv:1702.06838 [math.OC]} \BibitemShut
  {NoStop}%
\bibitem [{\citenamefont {Preskill}(2018)}]{Preskill_2018}%
  \BibitemOpen
  \bibfield  {author} {\bibinfo {author} {\bibfnamefont {J.}~\bibnamefont
  {Preskill}},\ }\href {\doibase 10.22331/q-2018-08-06-79} {\bibfield
  {journal} {\bibinfo  {journal} {Quantum}\ }\textbf {\bibinfo {volume} {2}},\
  \bibinfo {pages} {79} (\bibinfo {year} {2018})}\BibitemShut {NoStop}%
\bibitem [{\citenamefont {{Brandao}}\ and\ \citenamefont
  {{Svore}}(2017)}]{Brandao2017quantum}%
  \BibitemOpen
  \bibfield  {author} {\bibinfo {author} {\bibfnamefont {F.~G. S.~L.}\
  \bibnamefont {{Brandao}}}\ and\ \bibinfo {author} {\bibfnamefont {K.~M.}\
  \bibnamefont {{Svore}}},\ }in\ \href {\doibase 10.1109/FOCS.2017.45} {\emph
  {\bibinfo {booktitle} {2017 IEEE 58th Annual Symposium on Foundations of
  Computer Science (FOCS)}}}\ (\bibinfo {year} {2017})\ pp.\ \bibinfo {pages}
  {415--426}\BibitemShut {NoStop}%
\bibitem [{\citenamefont {{Van Apeldoorn}}\ \emph {et~al.}(2017)\citenamefont
  {{Van Apeldoorn}}, \citenamefont {{Gilyén}}, \citenamefont {{Gribling}},\
  and\ \citenamefont {{de Wolf}}}]{VanApeldoorn2017quantum}%
  \BibitemOpen
  \bibfield  {author} {\bibinfo {author} {\bibfnamefont {J.}~\bibnamefont {{Van
  Apeldoorn}}}, \bibinfo {author} {\bibfnamefont {A.}~\bibnamefont
  {{Gilyén}}}, \bibinfo {author} {\bibfnamefont {S.}~\bibnamefont
  {{Gribling}}}, \ and\ \bibinfo {author} {\bibfnamefont {R.}~\bibnamefont {{de
  Wolf}}},\ }in\ \href {\doibase 10.1109/FOCS.2017.44} {\emph {\bibinfo
  {booktitle} {2017 IEEE 58th Annual Symposium on Foundations of Computer
  Science (FOCS)}}}\ (\bibinfo {year} {2017})\ pp.\ \bibinfo {pages}
  {403--414}\BibitemShut {NoStop}%
\bibitem [{\citenamefont {Brandão}\ \emph {et~al.}(2019)\citenamefont
  {Brandão}, \citenamefont {Kueng},\ and\ \citenamefont
  {França}}]{Brandao2019faster}%
  \BibitemOpen
  \bibfield  {author} {\bibinfo {author} {\bibfnamefont {F.~G. S.~L.}\
  \bibnamefont {Brandão}}, \bibinfo {author} {\bibfnamefont {R.}~\bibnamefont
  {Kueng}}, \ and\ \bibinfo {author} {\bibfnamefont {D.~S.}\ \bibnamefont
  {França}},\ }\href@noop {} {\enquote {\bibinfo {title} {Faster quantum and
  classical sdp approximations for quadratic binary optimization},}\ }
  (\bibinfo {year} {2019}),\ \Eprint {http://arxiv.org/abs/1909.04613}
  {arXiv:1909.04613 [cs.DS]} \BibitemShut {NoStop}%
\bibitem [{\citenamefont {Grant}\ and\ \citenamefont {Boyd}(2013)}]{CVX}%
  \BibitemOpen
  \bibfield  {author} {\bibinfo {author} {\bibfnamefont {M.}~\bibnamefont
  {Grant}}\ and\ \bibinfo {author} {\bibfnamefont {S.}~\bibnamefont {Boyd}},\
  }\href {http://cvxr.com/cvx} {\bibfield  {journal} {\bibinfo  {journal}
  {\mbox{CVX}: Matlab software for disciplined convex programming, version 2.0
  beta}\ } (\bibinfo {year} {2013})}\BibitemShut {NoStop}%
\bibitem [{\citenamefont {ApS}(2019)}]{mosek}%
  \BibitemOpen
  \bibfield  {author} {\bibinfo {author} {\bibfnamefont {M.}~\bibnamefont
  {ApS}},\ }\href {http://docs.mosek.com/9.0/toolbox/index.html} {\emph
  {\bibinfo {title} {The MOSEK optimization toolbox for MATLAB manual. Version
  9.0.}}} (\bibinfo {year} {2019})\BibitemShut {NoStop}%
\end{thebibliography}%

\end{document}